\documentclass[useAMS,usenatbib]{mn2e}
\usepackage{graphicx}
\usepackage{amsmath,amssymb,wasysym}
\usepackage{color}
\pdfoutput=1
\graphicspath{{figures/}}

\usepackage{hyperref}
\providecommand{\eprint}[1]{preprint (arXiv:\href{http://arxiv.org/abs/#1}{#1})}
\providecommand{\adsurl}[1]{\href{#1}{ADS}}

\def\msun{M$_{\astrosun}$}
\def\msunsp{M$_{\astrosun}\;$}

\title[I. The impact of r-process nucleosynthesis] {
The long-term evolution of neutron star merger remnants --
I. The impact of r-process nucleosynthesis}
\author[Rosswog et
al.]{S. Rosswog$^{1}$\thanks{E-mail:
    Stephan.Rosswog@astro.su.se}, O. Korobkin$^{1}$, A. Arcones$^{2,3}$, 
    F.-K. Thielemann$^{4}$, T. Piran$^{5}$\\
  $^{1}$The Oskar Klein Centre, Department of Astronomy, AlbaNova, 
        Stockholm University, SE-106 91 Stockholm, Sweden\\
  $^{2}$Institut f{\"u}r Kernphysik, Technische Universit{\"a}t Darmstadt, 64289 Darmstadt, Germany\\
  $^{3}$GSI Helmholtzzentrum f\"ur Schwerionenforschung, Planckstr. 1, 64291 Darmstadt, Germany\\
  $^{4}$Department of Physics, University of Basel, Klingelbergstrasse 82, CH-4056 Basel, Switzerland\\
  $^{5}$Racah Institute of Physics, The Hebrew University, Jerusalem 91904, Israel\\
}

\begin{document}

\date{Accepted 2013. Received 2013; in original form 2013}

\pagerange{\pageref{firstpage}--\pageref{lastpage}} \pubyear{2013}
\setlength\parindent{0pt}

\maketitle

\label{firstpage}

\begin{abstract}
We follow the longterm evolution of the dynamic ejecta of neutron star mergers
for up to 100 years and over a density range of roughly 40 orders of magnitude.
We include the nuclear energy input from the freshly synthesized, radioactively
decaying nuclei in our simulations and study its effects on the remnant
dynamics.  Although the nuclear heating substantially alters the longterm
evolution, we find that running nuclear networks over purely hydrodynamic
simulations (i.e.  without heating) yields actually acceptable nucleosynthesis
results.  The main dynamic effect of the radioactive heating is to quickly
smooth out inhomogeneities in the initial mass distribution, subsequently the
evolution proceeds self-similarly and after 100 years the remnant still carries
the memory of the initial binary mass ratio. We also explore the nucleosynthetic
yields for two mass ejection channels. The dynamic ejecta very robustly produce
`strong' r-process elements with $A>130$ with a pattern that is essentially
independent of the details of the merging system. From a simple model we find
that neutrino-driven winds yield `weak' r-process contributions with $50 < A <
130$ whose abundance patterns vary substantially between different merger cases.
This is because their electron fraction, set by the ratio of neutrino
luminosities, varies considerably from case to case. Such winds do not produce
any $^{56}$Ni, but a range of radioactive isotopes that are long-lived enough to
produce a second, radioactively powered electromagnetic transient in addition to
the `macronova' from the dynamic ejecta.
While our wind model is very simple, it nevertheless demonstrates the
potential of such neutrino-driven winds for electromagnetic transients and it
motivates further, more detailed neutrino-hydrodynamic studies.
The properties of the mentioned transients are discussed in more detail in a companion paper.
\end{abstract}


\section{Introduction}

The importance of compact binary mergers as sources of gravitational waves
\citep[GW; e.g.][]{cutler94,maggiore08} and as central engines of short Gamma-ray bursts 
\citep{eichler89,narayan91,piran04,nakar07,lee07} has been appreciated for
decades. Maybe more surprising is the fact that compact binary mergers have 
only been considered {\em seriously} as a possible production site of rapid neutron capture
(`r-process') elements in roughly the last decade. This is particularly
astonishing since there is general agreement that the final fate after their inspiral 
is a violent event in an extremely neutron-rich environment, exactly what 
is recognized to be needed as a cauldron to forge r-process elements. Moreover, 
\cite{lattimer74} had pointed out this possibility and they discussed its
impact on cosmic nucleosynthesis even before the first binary neutron star had 
been discovered. 
In \cite{lattimer77} r-process nucleosynthesis is discussed in the
context of a cold decompression of neutron star matter.
In the last decade a number of studies
found core-collapse supernovae to be seriously challenged in providing suitable conditions
to produce heavy r-process elements 
\citep[e.g.][]{arcones07,fischer10,huedepohl10,roberts10}\footnote{A possible exception may be stars 
with the particular, and probably rare, combination of fast rotation and very
large pre-collapse magnetic field \citep{winteler12b}.
}. 
In parallel, several studies found that a robust r-process 
occurs in the dynamic ejecta of neutron star mergers 
\citep{freiburghaus99b,goriely11a,roberts11,korobkin12a}. These developments have changed the general
perceptions and nowadays, neutron stars are accepted at least as serious contenders of 
core-collapse supernovae if not even {\em the} major candidate sites for the production 
of the heaviest, `strong r-process' elements ($A \ga 130$). Even if the latter is true,
at least one but probably more r-process sites must exist in addition. The dynamic ejecta of 
neutron star mergers are probably not a source of `weak r-process' \citep{sneden08}, 
but the neutrino-driven winds that emerge in the aftermath of a merger  \citep{dessart09} 
could plausibly contribute to this r-process component. A third process by which
neutron-rich matter may be shed in substantial amounts is the final disintegration of
an accretion disc that occurs after many viscous time-scales ($\sim$ seconds)
when viscous dissipation and the recombination of nucleons into  light nuclei
conspire to unbind a substantial fraction of whatever is left from the disc at that
time \citep{lee07,beloborodov08,metzger08,metzger09b,lee09}. Recent work \citep{fernandez13}
suggests that $\sim 10^{-2}$ \msunsp become unbound with $Y_e\sim0.2$, in addition to the 
dynamic ejecta and the wind material. Nucleosynthesis features of such disc ejecta 
(partially also invoking neutrino flavour oscillations) have been 
studied by \cite{surman08}, \cite{caballero12}, \cite{wanajo12} and \cite{malkus12}.
\\
The astrophysical relevance of the dynamic ejecta from compact binary mergers, however,
goes beyond their roles as promising nucleosynthesis site. They are believed to also trigger 
short-lived electromagnetic transients powered by radioactivity, so-called `macronovae'
\citep{li98,kulkarni05,rosswog05a,metzger10b, roberts11,goriely11a,metzger12a,kelley12,rosswog13a,piran13a,barnes13a}. 
As we will discuss below, also the neutrino-driven winds may actually produce radioactively
powered transients, though with different nuclear and electromagnetic properties than those from
the dynamic ejecta.
Both the LIGO and Virgo GW detectors 
are currently being upgraded \citep{abbott09a,smith09,sengupta10} to a more than 10 times better 
sensitivity than the original versions of the instruments. This will increase the volume of 
accessible astrophysical sources by more than a factor of 1000 reaching a detection horizon of 
a few hundred Mpc for nsns mergers and about a Gpc for nsbh mergers \citep{abadie10}. Since
the first events are expected to be around or even below threshold, accompanying electromagnetic 
signatures, such as r-process powered macronovae, could substantially boost the confidence in such
a marginal detection \citep{nissanke13,kasliwal13}. The heating from the nuclear
decay may also suspend fallback accretion \citep{metzger10a} and thus cause a gap between 
prompt and late-time activity in short GRBs.
Months after the merger, the dynamic ejecta may trigger radio flares when they
dissipate their kinetic energies in the ambient medium \citep{nakar11a,piran13a} and this
could possibly provide lower limits on the merger rates 
even before the GW detectors are fully operational.
\\
In this paper we present a new study of neutron star merger ejecta that goes beyond previous work
in several ways. First, rather than performing hydrodynamic simulations for $\sim 20$ ms and then 
extrapolating the thermodynamic ejecta histories to $\sim 10$ s 
\citep{freiburghaus99b,goriely11a,roberts11,korobkin12a}, our simulations directly follow the 
hydrodynamic evolution of the ejecta for time-scales as long as 100 years after the merger. 
We stop our simulations at this point since, by this time, the ejecta likely
have swept up an amount of mass from the ambient medium that is comparable to their own 
mass and so they start to slow down. Where and when exactly this will happen depends sensitively
on the environment in which the merger occurs. This, in turn, depends on the kick velocity and the 
inspiral time of the particular binary system. The ejecta from mergers that occur in the galactic 
plane may be braked much earlier than those of mergers occuring a few kpc outside of their host 
galaxies, see the estimates below.\\
\begin{figure*}
 \centerline{
   \includegraphics[width=0.5\textwidth]{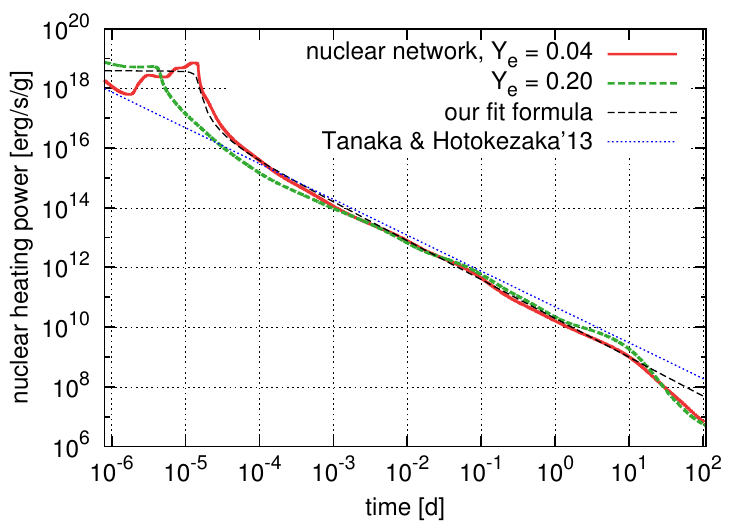} 
   \includegraphics[width=0.5\textwidth]{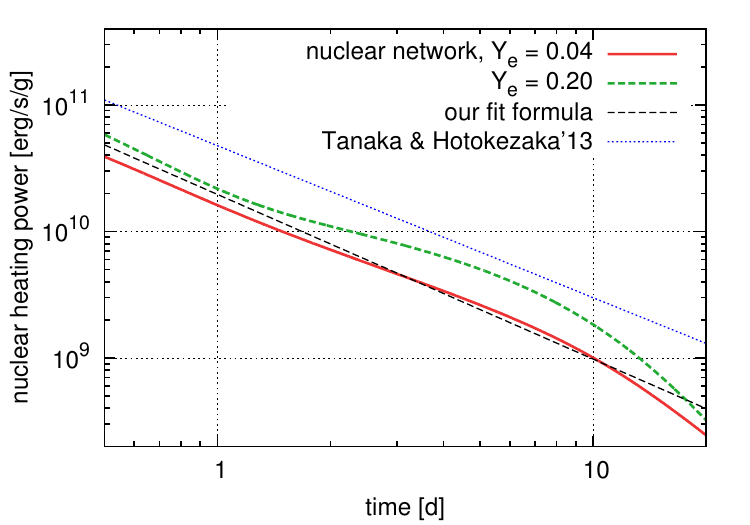} 
 }
 \caption{ 
    A comparison of the nuclear heating rates used in different 
    approaches on different time-scales. The left panel depicts the whole 
    duration of the event while the right panel focuses on a time scale of 
    a few days, which is most relevant for the macronovae emission.
    The solid (red) and dashed (green) lines display heating rates for initial
    $Y_e=0.04$ (the typical value that results from initial beta-equilibrium plus subsequent
    electron/positron captures in our simulations) and $Y_e=0.2$ (the value adopted for dynamical ejecta in 
    \protect\cite{barnes13a}, based on the work of \protect\cite{roberts11}).
    Also displayed is our analytic fit from~\protect\cite{korobkin12a} and the
    one used in~\protect\cite{tanaka13a}. For proper comparison, we have 
    set the heating efficiency parameter $\epsilon_{\rm th}=1$. Note that the 
    increased $Y_e$ value results in a substantially larger heating rate in the
    epoch that is most relevant for macronovae.
 }
\label{fig:compare_edot}
\end{figure*}
\begin{figure*} 
   \centerline{\includegraphics[width=0.9\textwidth]{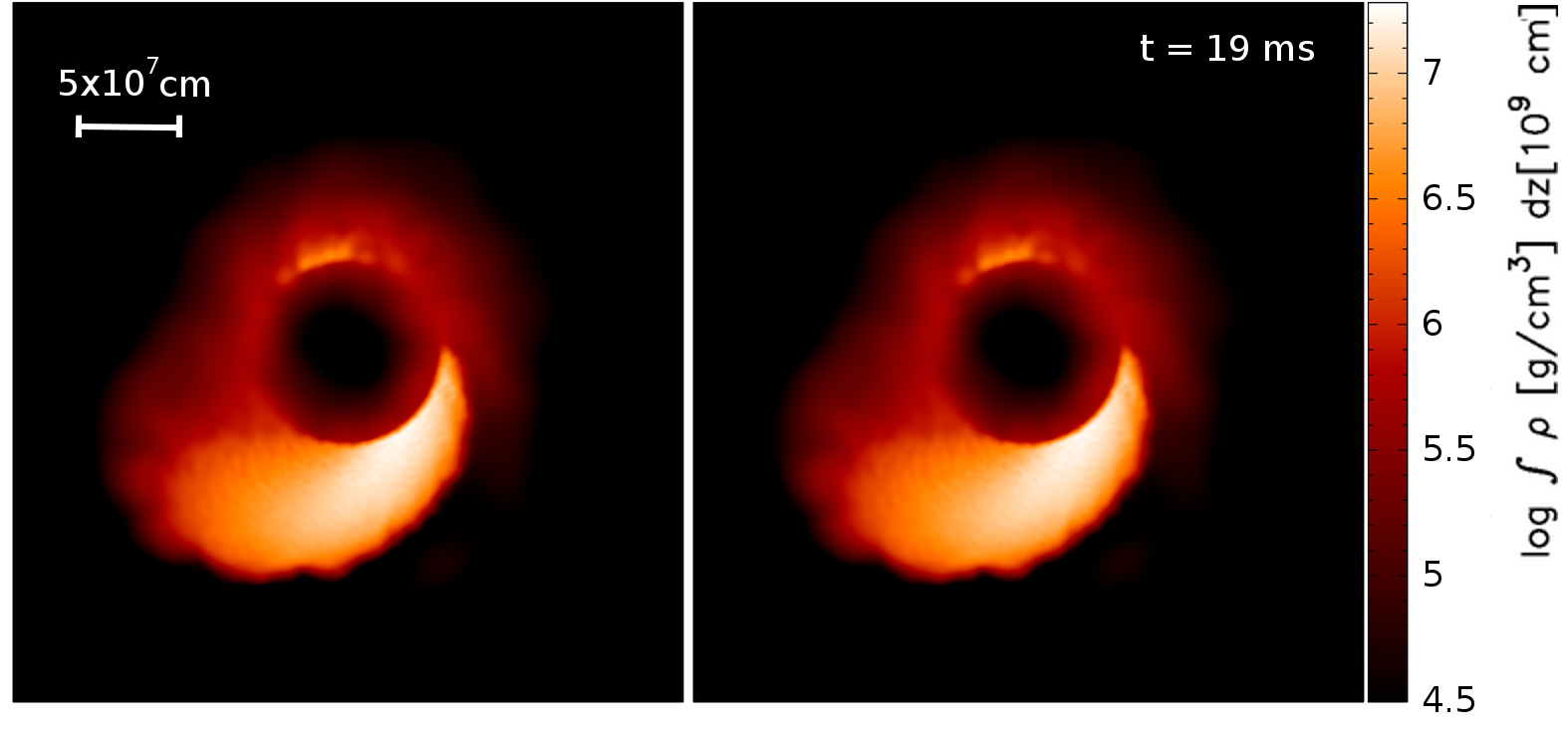}}
   \centerline{\includegraphics[width=0.9\textwidth]{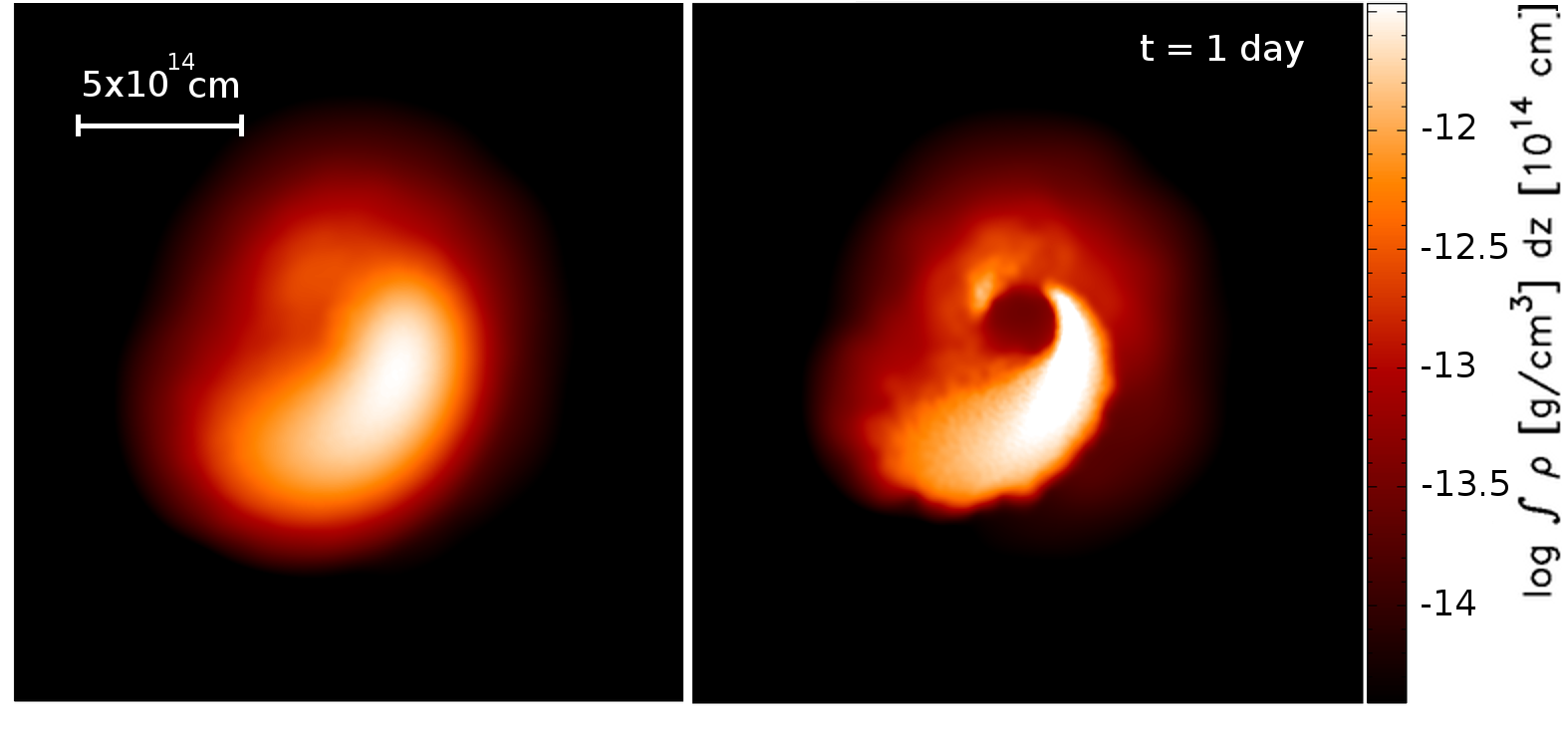}}
   \centerline{\includegraphics[width=0.9\textwidth]{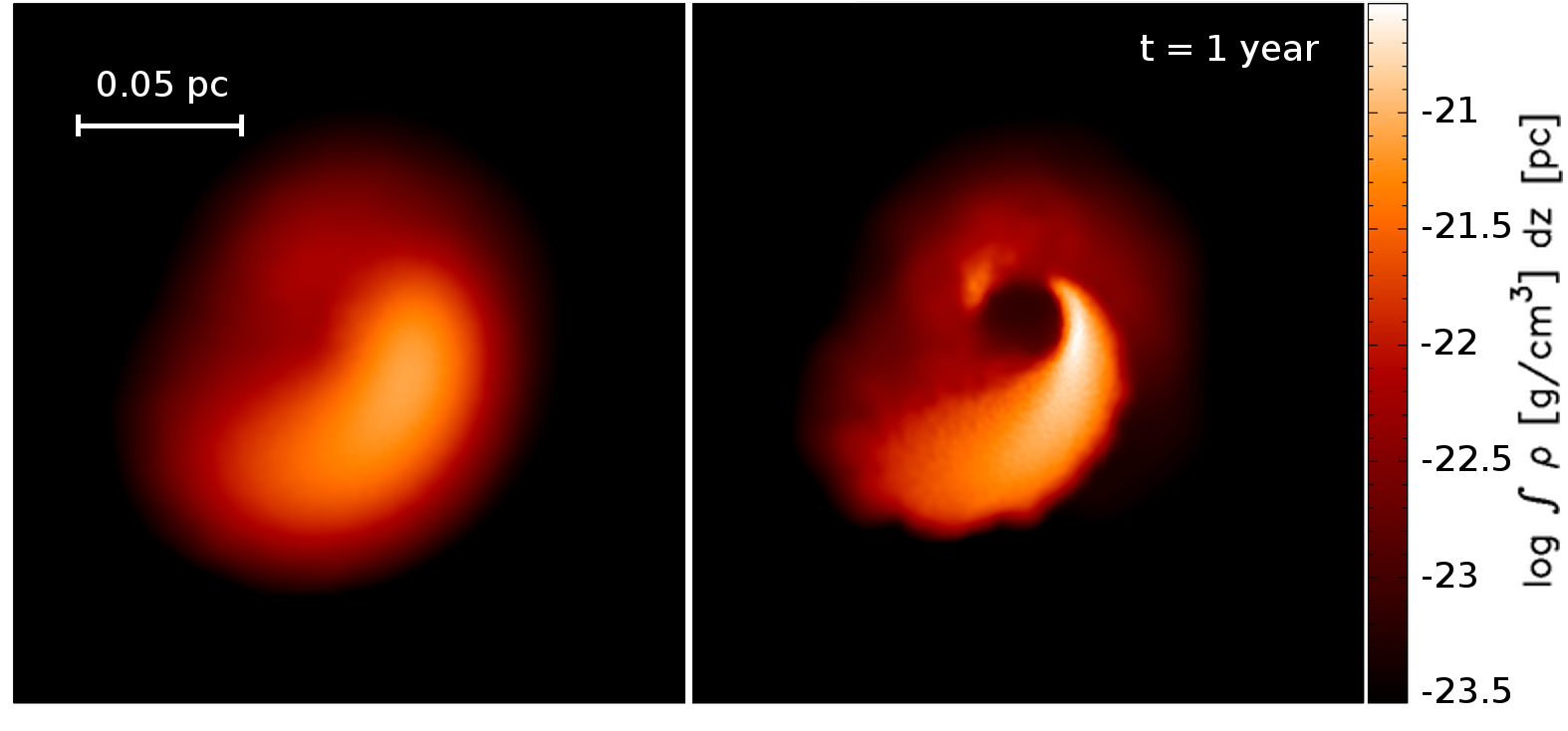}}
   \caption{Illustration of the effect of radioactive heating on the remnant evolution
               (for run B, see Tab.~\ref{tab:runs}; 1.3 and 1.4 \msun). The evolution including the effects 
                from radioactive decays is shown in the left column, and the one without in the right column. 
                The first row shows the initial configuration (corresponding to $t= 19$ ms of 
                our original merger simulation; in the spherical central region 
                matter has been cut out and has been replaced by a point mass). The second row corresponds
                to the matter configuration after $t= 1$ d, roughly when the resulting `macronova' will reach its peak
                bolometric luminosity. The last row shows the remnant structure at $t= 1$ year.}
   \label{fig:heating_vs_no_heating1}
\end{figure*}
\begin{figure} 
   \centerline{
     \includegraphics[width=0.56\textwidth]{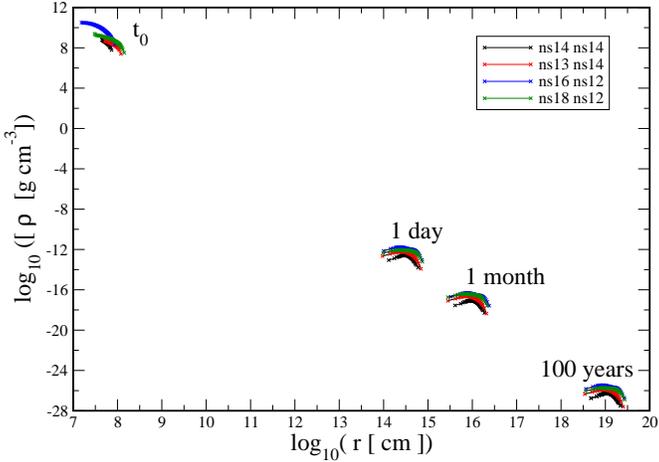}} 
      \caption{Density distribution (50 bins for each case) inside the remnants at chosen snapshots. 
               For each time label, the closest available data set was used.
               Depending on the exact merger environment, the late-time
               evolution may be impacted by ambient medium effects.}
   \label{fig:density_evolution}
\end{figure}
Secondly, all existing studies have assumed that the energetic 
feedback from nuclear reactions on the hydrodynamic evolution can be 
neglected, and that it is safe to just post-process given trajectories for nucleosynthesis. 
The standard approach is to ignore nuclear energy release in the hydrodynamics part of the calculation,
subsequently run  nuclear networks along given density trajectories and finally to reconstruct 
the temperature histories from the entropy that was generated in nuclear reactions \citep{freiburghaus99b}. 
In other words, it is usually assumed that the nuclear energy input does not substantially 
impact on the density evolution. 
Analytic one-zone models~\citep{goriely11b} seem to support this assumption, but as
we will show below, the nuclear feedback has indeed a serious impact on the longterm evolution 
of neutron star merger remnants. Nevertheless, on the short time-scales on which the r-process
occurs ($\sim$ seconds) its impact is still small enough so that post-processing hydrodynamic 
trajectories is an admissible procedure which yields acceptable nucleosynthesis results.\\
Third, we are able to follow the ejecta through the times when radioactively powered transients
should peak ($\sim$ days) and up to the point when the radio flares 
from the interaction with the ambient medium are expected ($\sim$ years). The available
geometric information allows us to abandon the assumption of spherical symmetry 
that was adopted in all previous calculations of `macronovae' light curves. 
This complements recent approaches by \cite{barnes13a} and \cite{kasen13a} who use 
sophisticated radiative transfer, but make strongly simplifying assumptions about the 
ejecta geometry. These results are discussed in a companion paper \citep{grossman13a},
subsequently referred to as `Paper II'.\\
This paper is structured as follows. In Sec.~\ref{sec:sims} we briefly summarize our simulations
and in Sec.~\ref{sec:remnant} we discuss the remnant structure 
and its close-to-homologous evolution during the first century after the merger where we pay particular attention to the effects 
of radioactive heating. 
In Sec.~\ref{sec:nucleo} the nucleosynthesis from both dynamic ejecta and 
neutrino-driven winds is discussed, the corresponding radioactively powered electromagnetic signals 
are examined in detail in Paper II. We summarize our results in Sec. \ref{sec:discussion}.

\section{Simulations}
\label{sec:sims}

\subsection{When to stop?}
In our simulations we follow the ejecta of neutron star mergers as they expand into vacuum. In Nature, 
a merger is engulfed by an external, though dilute medium of a density $\rho_{\rm amb}= m_{\rm p} n_{\rm amb}$
that depends on the actual merger location. Mergers that occur early after the neutron star binaries
have formed take place close to the midplane of their host galaxies where the density may be $n \sim 1$ cm$^{-3}$.
On the other hand, binary systems that had time to travel a few kpc out of the midplane 
\citep{narayan92,fryer99a,bloom02,rosswog03c,fong10}
may occur in a much lower density surrounding. The initial 
expansion stages will be unaffected by the ambient medium, but once the
swept up amount of matter is 
comparable to the ejected mass, $m_{\rm su} \approx m_{\rm ej}$, the ejecta start to slow down. This 
transition defines the deceleration radius of
\begin{equation}
R_{\rm dec}= 0.5 \; {\rm pc} \left( \frac{m_{\rm ej}}{10^{-2} M_\odot} \;
                                 \frac{1 \: {\rm cm}^{-1/3}}{n_{\rm amb}} \right)^{1/3},
\label{eq:r_dec}
\end{equation}
which is reached after the deceleration time
\begin{equation}
\tau_{\rm dec}= 
15 \; {\rm yr} \; \left( \frac{m_{\rm ej}}{10^{-2} M_\odot} \;   
 \frac{1 \: {\rm cm}^{-1/3}}{n_{\rm amb}} \right)^{1/3}
\left( \frac{0.1 \: {\rm c}}{v_{\rm ej}} \right).
\label{eq:tau_dec}
\end{equation}
Both estimates are rather insensitive to the poorly known ambient matter density. Thus, ambient matter 
effects start to become noticeable after 15 (150) years in an environment of $n= 1 (10^{-3})$ cm$^{-3}$.
Since the effects from an ambient medium are not taken into account in our study, we stop the simulations
100 years after the coalescence.
\begin{table*}
\centering
 \begin{minipage}{140mm}
  \caption{Overview over the performed simulations: $m_1$ and $m_2$ are masses
  of the neutron stars, $N_{\rm SPH}$ is the number
  of SPH particles and $t_{\rm end}$ is the final simulation time.
  Properties of the dynamical ejecta include their mass $m_{\rm ej}$,
  kinetic energy $E_{\rm kin}$ and thermalized nuclear energy $E_{\rm nuc}$.
  Neutrino-related quantities: $L_{\nu_e}$ and $L_{\bar{\nu}_e}$ are
  neutrino and anti-neutrino luminosities, $\langle E\rangle_\nu$ and $\langle
  E\rangle_{\bar\nu}$ are the average neutrino energies, and $Y_e^{\rm fin, 
  wind}$ is the estimated asymptotic $Y_e$ in the neutrino driven wind
  \citep{qian96}.}
  \label{tab:runs}
\vspace*{0.3cm}
\begin{tabular}{@{}rccccccccccccccc@{}}
\hline
   Run   &  $m_1\quad m_2$ & $N_{\rm SPH}$ & $t_{\rm end}$ & $m_{\rm ej}$ &
   $E_{\rm kin}\quad E_{\rm nuc}$ & $L_{\nu_e}\quad L_{\bar{\nu}_e}$&
   $\langle E \rangle_{\nu_e}\quad\langle E \rangle_{\bar{\nu}_e}$  & $Y_e^{\rm fin, wind}$ &  \\
   \vspace{-0.2cm} \\
         &  (\msun) & $(10^6)$ & (ms) & ($10^{-2}$\msun) & ($10^{50}$ erg) & ($10^{52}$ erg/s) & (MeV) & (MeV) & &  \\
   \hline \\  
  A    & $1.4\quad1.4$ & 1.0   &  13.4  & 1.3 & $2.6\qquad0.8$ & $3.0\;\;\quad6.1 $& $7.8\qquad14.3$ &  0.28\\
  B    & $1.3\quad1.4$ & 2.7   &  20.3  & 1.4 & $2.4\qquad0.9$ & $3.1\;\;\quad6.0 $& $8.0\qquad14.4$ &  0.30\\  
  C    & $1.6\quad1.2$ & 1.0   &  14.8  & 3.3 & $6.8\qquad2.1$ & $7.0\;\;\;\;11.0$ & $9.5\qquad15.0$ &  0.36\\
  D    & $1.8\quad1.2$ & 1.0   &  21.4  & 3.4 & $7.5\qquad2.2$ & $5.0\;\;\quad7.2 $& $9.4\qquad13.6$ &  0.40\\
\end{tabular}
\end{minipage}
\end{table*}

\begin{figure*} 
   \centerline{
     \includegraphics[width=1.\textwidth]{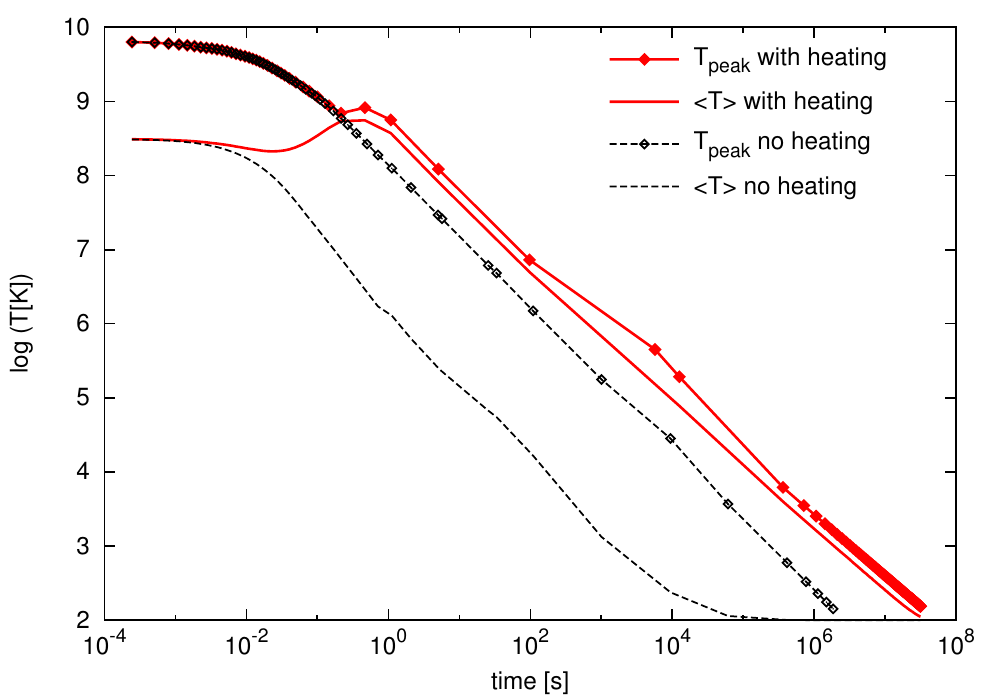} 
   }
      \caption{Illustration of the effect of radioactive heating on the temperature evolution
               (for run B, see Tab.~\ref{tab:runs}; 1.3 and 1.4 \msun). The peak temperature with diamond 
               symbol, and the average with a solid line; for comparison the values without heating are shown in black.}
   \label{fig:heating_vs_no_heating2_T}
\end{figure*}
\begin{figure*} 
   \centerline{
     \includegraphics[width=1.\textwidth]{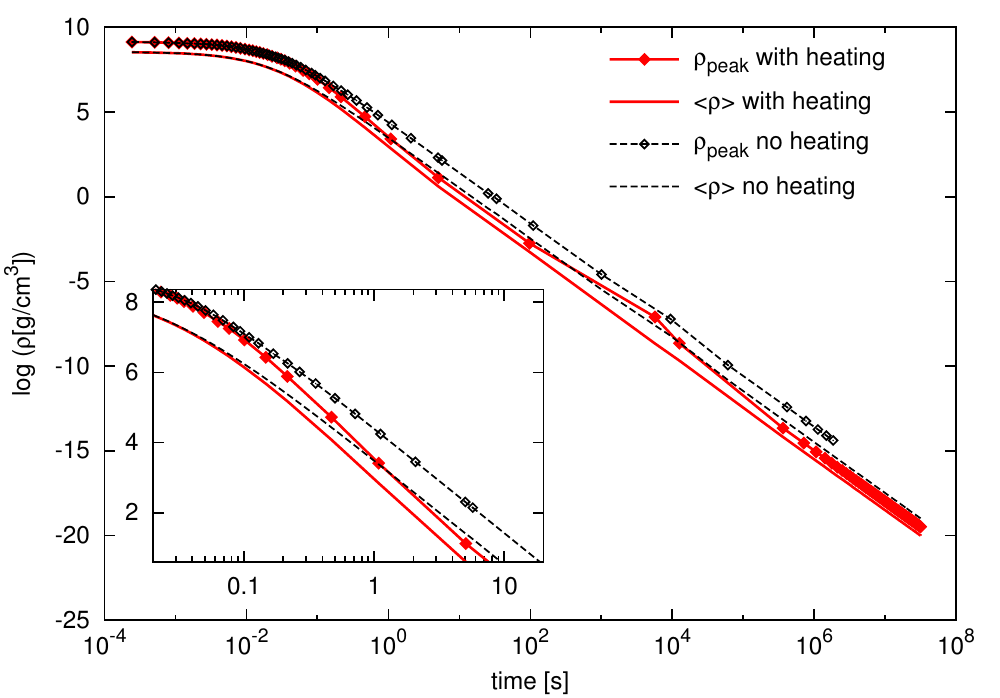} 
   }
      \caption{Illustration of the effect of radioactive heating 
               on the density evolution
               (for run B, see Tab.~\ref{tab:runs}; 1.3 and 1.4 \msun). 
               The labelling of the curves is as in the previous figures.
               Inset zooms on the density evolutions for the period around
               $t\sim1$~s, when heating has maximal impact on the density, 
               decreasing it by about one order of magnitude.
               }
   \label{fig:heating_vs_no_heating2_rho}
\end{figure*}

\subsection{Hydrodynamic simulations}
\label{sub:hydrosim}
Typically neutron star merger simulations are only performed up to  $\sim 20$ ms, mainly because
of serious time-step restrictions due to the Courant--Friedrichs--Lewy stability criterion.
The simulations that are presented here, in contrast, are performed up 
to 100 years after
the coalescence. Therefore, we have information on the remnant geometry throughout 
the phases where the major electromagnetic emission -- from radioactivity in the neutrino-driven winds
and dynamic ejecta and the interaction with ambient material-- is expected.\\
The simulations of this paper start from the final matter distributions of the runs
presented in \cite{rosswog13a} and \cite{rosswog13b}. 
These simulations were performed with a 3D, Smooth Particle Hydrodynamics (SPH) code whose
implementation details have been described in the literature \citep{rosswog00,rosswog02a,
rosswog03a,rosswog07c}. For a general overview over the SPH method, see, for example, the 
recent reviews of \cite{monaghan05}, \cite{rosswog09b} and \cite{springel10}.\\
The code that we use for the subsequent longterm evolution of the ejecta is a variant of the 
above described SPH code, but with different units and physics ingredients
\citep{rosswog08b}. Since the ejecta densities change in the first $\sim$ 100 years 
by as many as $\sim$40 orders of magnitude (!) special attention has to be paid to the 
equation of state (EOS). The initial merger simulations make use of the Shen-EOS \citep{shen98a,shen98b},
the longterm evolution is followed using the Helmholtz EOS \citep{timmes00a} 
which is the state-of-the-art for matter at roughly white dwarf densities. It uses in 
particular a completely general electron EOS which is recovered via a 
sophisticated interpolation scheme from pre-calculated tables. When the density and 
temperature drop during the matter expansion to values near the lower limits of the 
Helmholtz EOS, we smoothly switch over to a Maxwell--Boltzmann gas plus radiation.\\
Our main interest here is the long-term evolution of the dynamically ejected matter. As
stated above, this material cannot be evolved for long time-scales together with the dense 
inner parts of the remnant since the high sound speed near nuclear matter density ($\sim 0.3$ c) 
enforces prohibitively small numerical time-steps. Therefore, we replace the inner 
part of the remnant  at the end of the initial simulation ($\sim 18$ ms) by a point mass. 
This inner part is defined by a radius that safely includes all matter with close to zero 
and negative radial velocity, typically this radius is $R_{\rm cut}= 300$ km. Apart from 
reducing the SPH particle number, this configuration now allows for much larger (and increasing!)
numerical time-steps which make this long-term simulation feasible in the first place.\\
We follow the evolution of the ejecta including the radioactive heating for a number of
exemplary systems: a) an equal mass merger with $2 \times 1.4$ \msun, b) a merger with a slight
asymmetry, 1.4 and 1.3 \msun, c) a merger of a 1.6-1.2 \msunsp system and finally d) the merger of
a 1.8 \msunsp ns with a 1.2 \msunsp ns (see Tab.~\ref{tab:runs}). For all systems the simulations stop
100 years after the merger. \\
For the involved nucleosynthesis calculations we make use of the nuclear reaction network of Winteler
\citep{winteler12,winteler12b} which represents an update of the BasNet network \citep{thielemann11}.

\subsection{Implementation of the r-process heating}
During the hydrodynamical evolution we include the heating due to radioactive
decays. We had recently explored the nucleosynthesis in neutron star
merger ejecta \citep{korobkin12a} and found that, 
in agreement with the findings of other groups
\citep{metzger10b,roberts11,goriely11a},
the heating history is rather insensitive to details of both the merging system and 
the individual matter trajectory can be well-fit as a function of time. 
In the current study, we use
fit formulae for the radioactive energy input, $\dot{\epsilon}_{\rm nuc}$, and for the average
nucleon and proton number, $\bar{A}$ and $\bar{Z}$, that are needed
to call the EOS. The expressions that we use in this study
are provided in Appendix A. 
The energy produced by the r-process comes mainly from beta decays \citep{metzger10b}. 
Initially, the r-process path stays far from stability due to the high neutron densities. 
During this phase, the neutron separation energy for the nuclei in the r-process path is 
$S_n \sim 2 - 3$~MeV, which is significantly smaller than the typical beta-decay Q-values 
around 10~MeV.
Fission can be very important in neutron 
star mergers \citep{freiburghaus99b}; however it provides significantly less energy 
than the beta decays \citep{metzger10b}. Therefore, the energy generation is dominated by
beta decays and initially stays approximately constant because the high neutron 
density and fission cycling maintain the matter at a given path far from stability 
where all beta decay rates are similar. After the neutrons have run out the r-process matter 
decays to stability and the contribution of beta decays to the energy generation is proportional
to $t^{-\alpha}$ \citep{metzger10b,korobkin12a}. For the phase of constant energy generation we 
have tried different values of $\dot{\epsilon}_{\mathrm{nuc}}$ in post-processing calculations 
and find that the evolution of temperature does not strongly vary for rates between 2 and 
8~MeV nuc$^{-1}$s$^{-1}$. In addition, only an upper limit can be 
determined for the energy that contributes to heat the matter because an unknown part of it 
escapes in the form of electron antineutrinos.\\
To account for neutrino energy losses associated with $\beta$-decays, we introduce 
a heating efficiency parameter $\epsilon_{\rm th}$  which denotes the fraction of 
nuclear power which is retained in the matter. \cite{metzger10b} argue that 
this fraction must be $\epsilon_{\rm th} \approx 0.25 ... 1$. Here and in Paper II we
adopt a value of $\epsilon_{\rm th} = 0.5$.\\

While details may benefit from an even more accurate heating treatment,
we consider our prescription accurate enough for this first study to explore whether and 
where heating by radioactive decay makes a noticeable difference with respect to the 
purely hydrodynamic evolution. We include the heating term explicitly in our energy 
equation and since the typical heating time-scales are similar to the hydrodynamic
time-step, we can use a single time integration scheme for the whole system of equations. This 
is different from the case where a nuclear network is coupled to the hydrodynamics. 
The latter case usually requires operator splitting techniques (e.g. Sec. 2 
in \cite{rosswog09a}), since the nuclear and hydrodynamic time-scales can differ by many orders
of magnitude.

\begin{figure*} 
   \centerline{ \includegraphics[height=3.7cm]{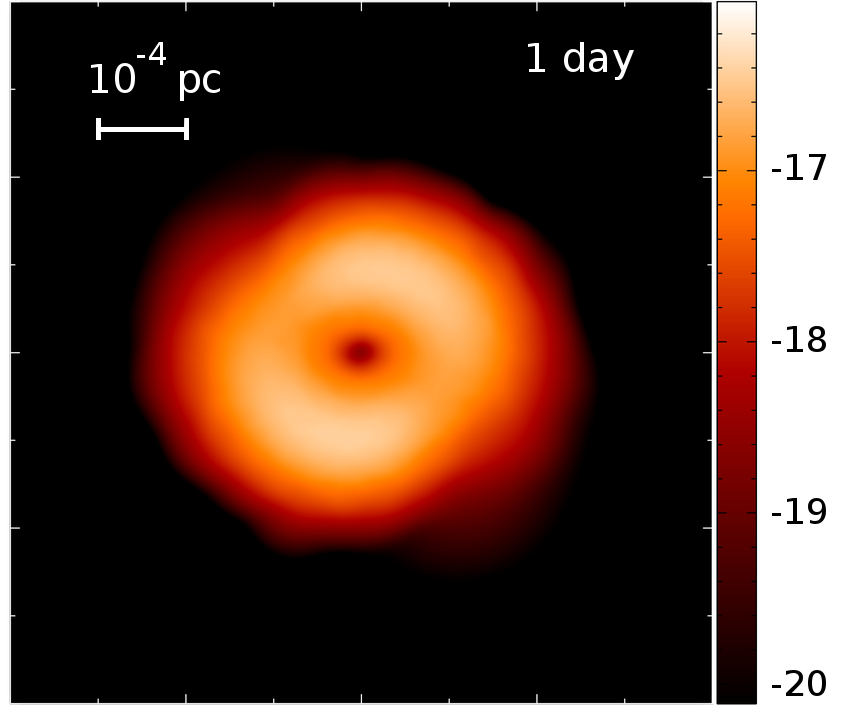} 
                \includegraphics[height=3.7cm]{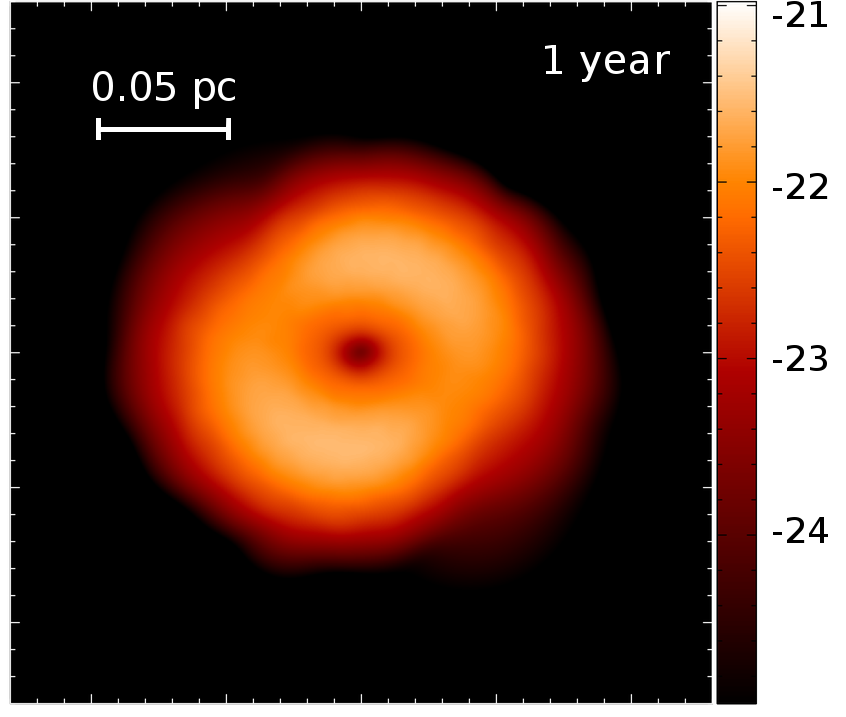} 
                \includegraphics[height=3.7cm]{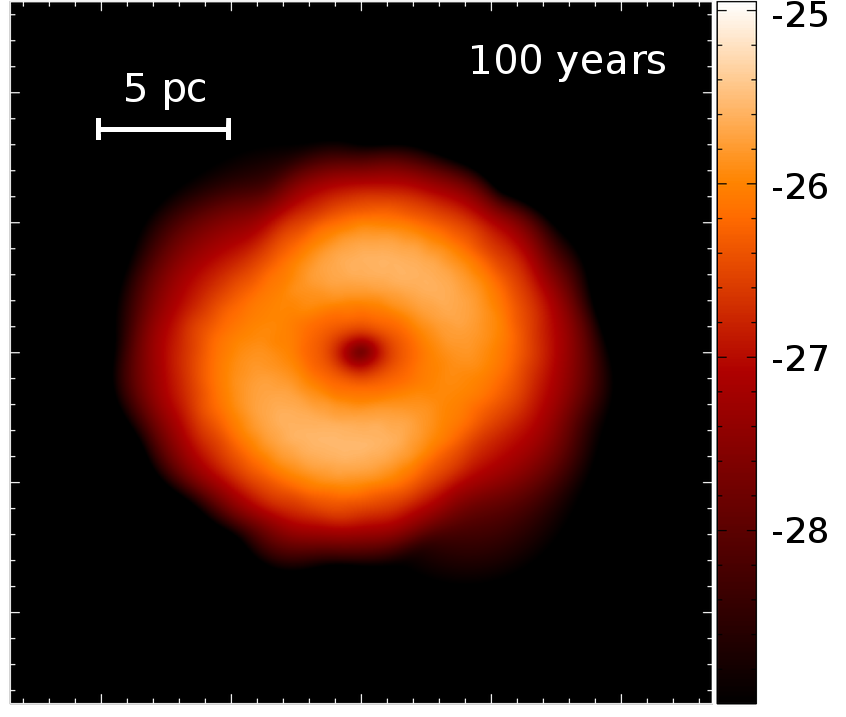}
                \includegraphics[height=3.7cm]{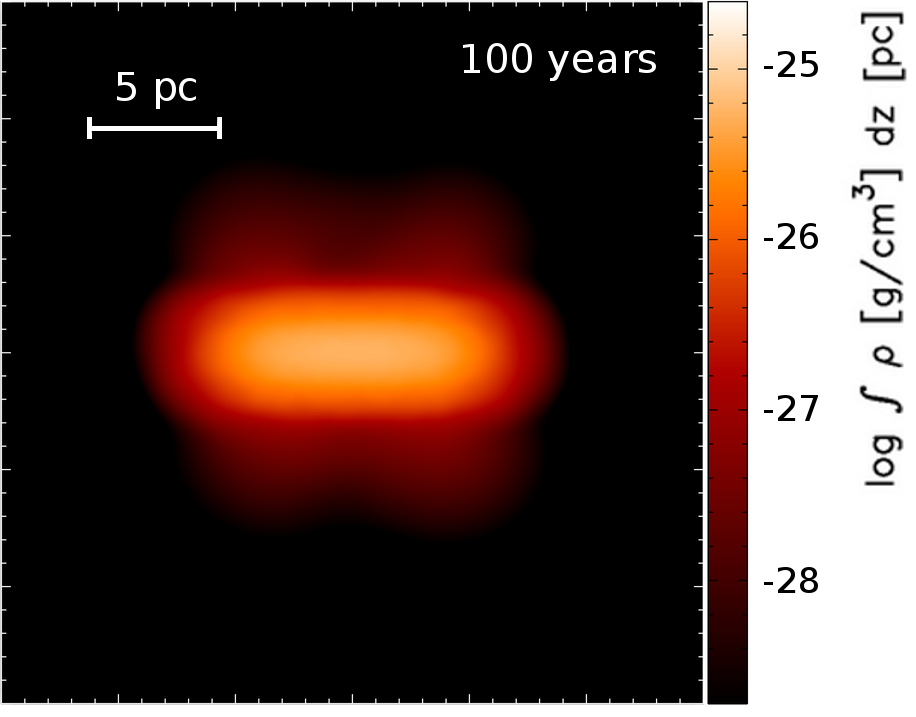}       
              }
   \centerline{ \includegraphics[height=3.7cm]{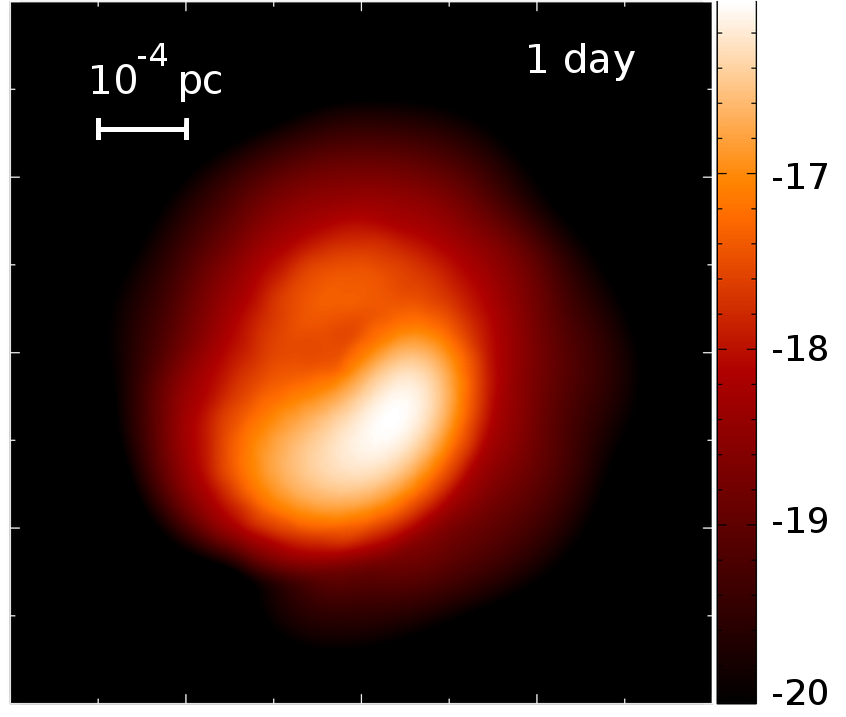} 
                \includegraphics[height=3.7cm]{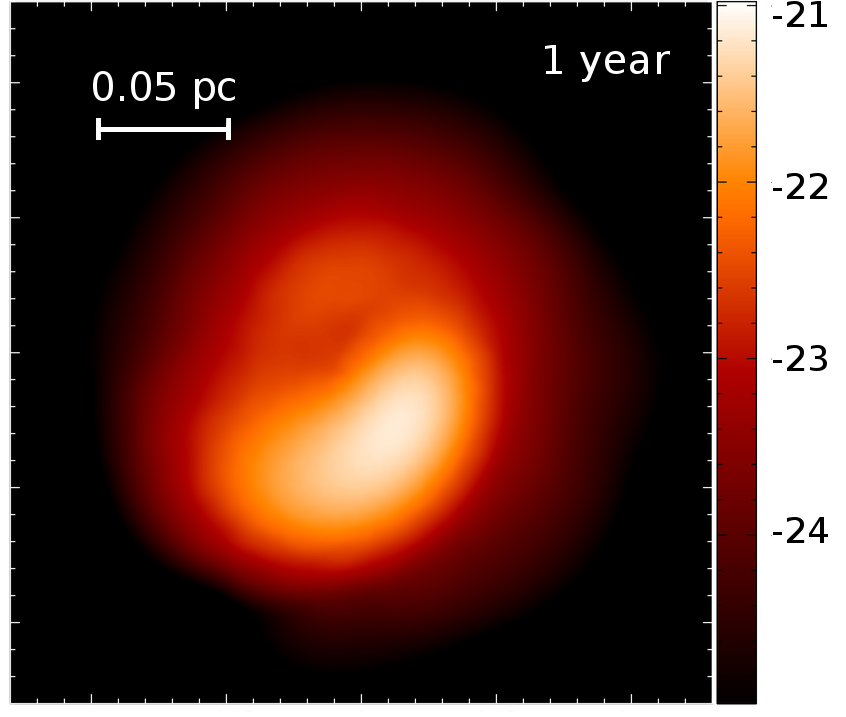} 
                \includegraphics[height=3.7cm]{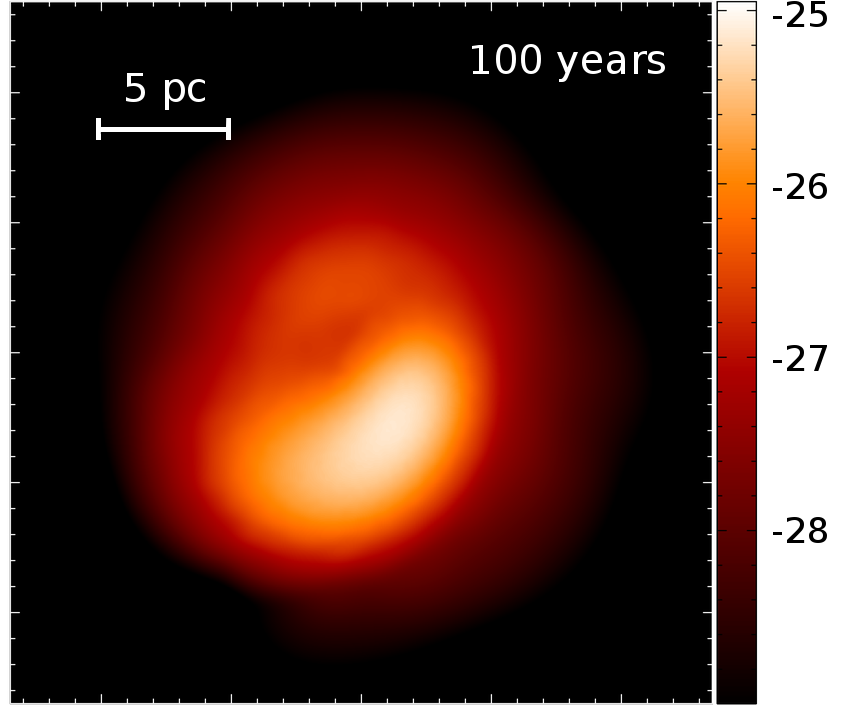}
                \includegraphics[height=3.7cm]{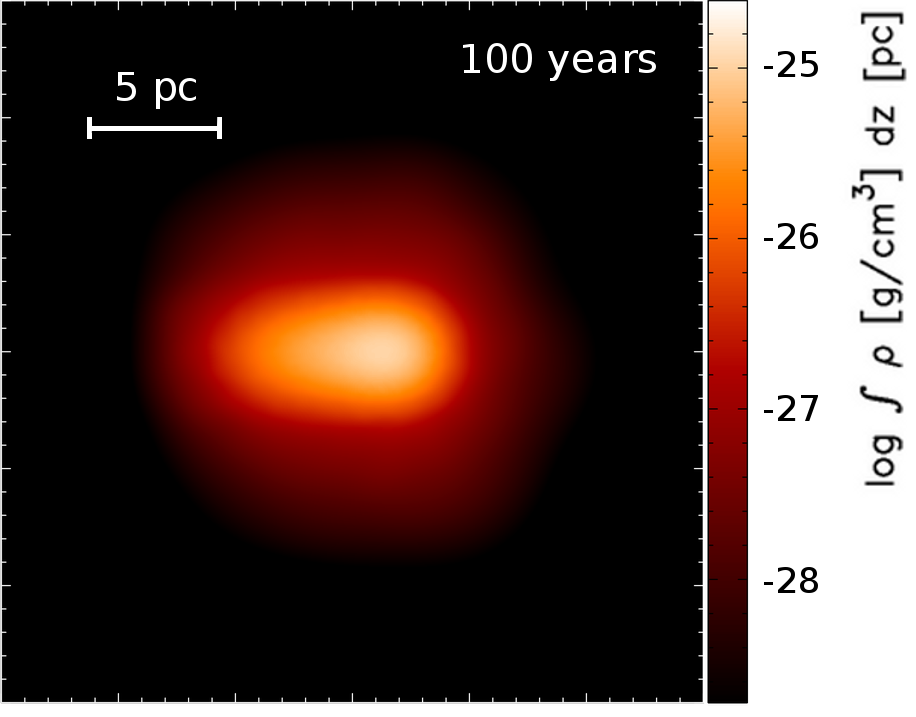}
              }
   \centerline{ \includegraphics[height=3.7cm]{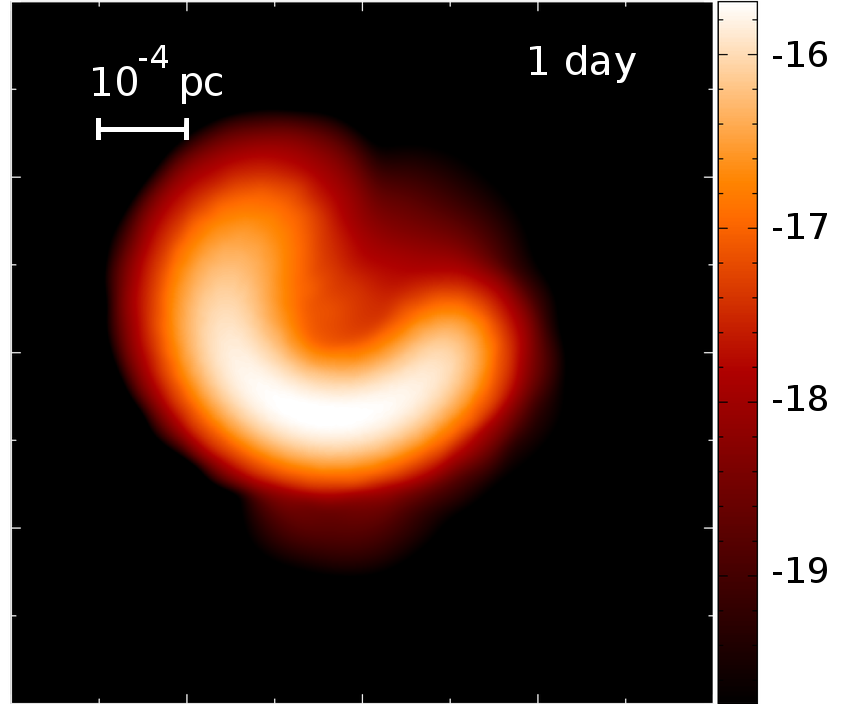} 
                \includegraphics[height=3.7cm]{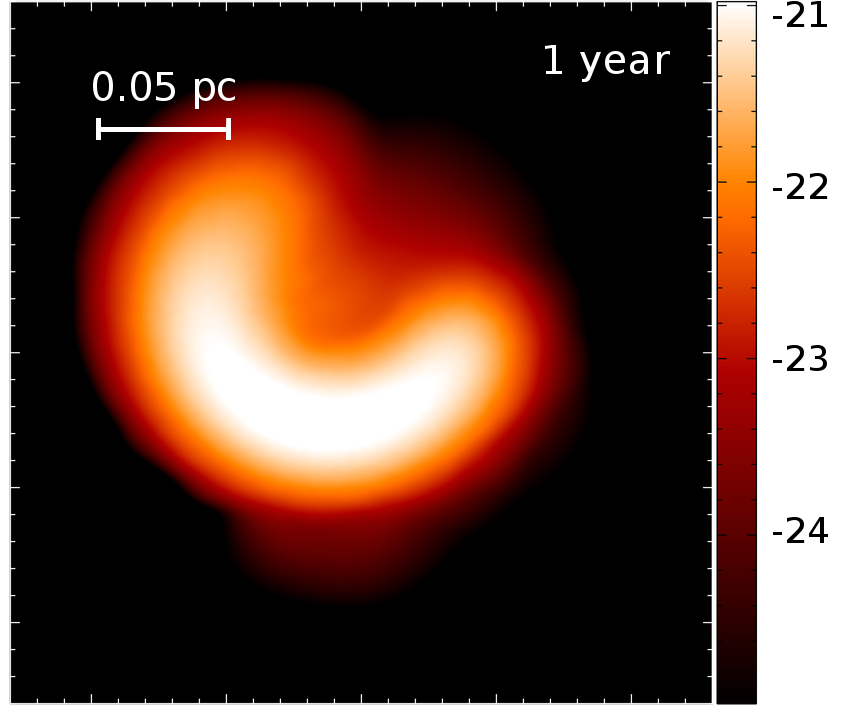} 
                \includegraphics[height=3.7cm]{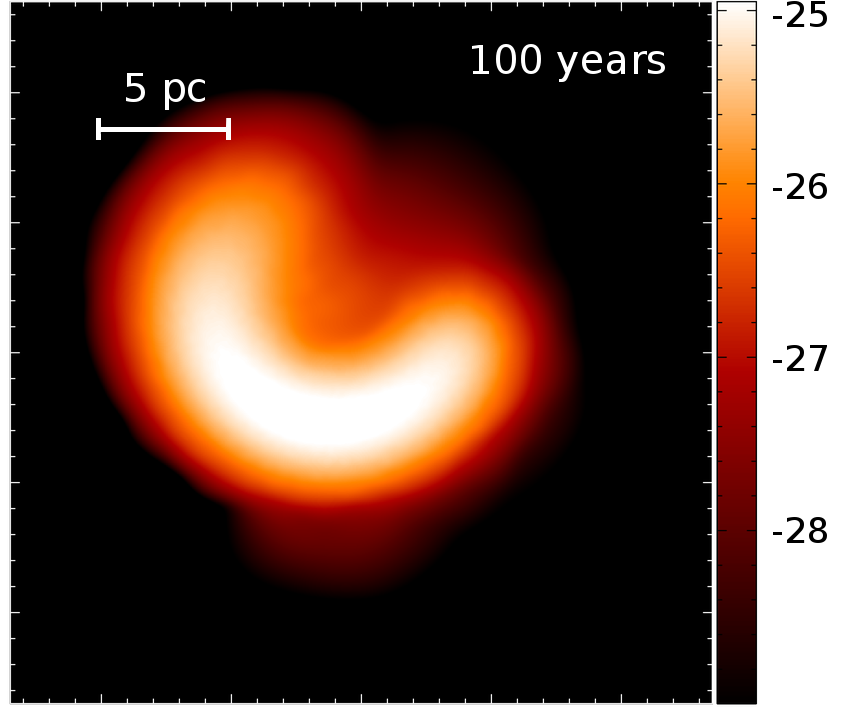}
                \includegraphics[height=3.7cm]{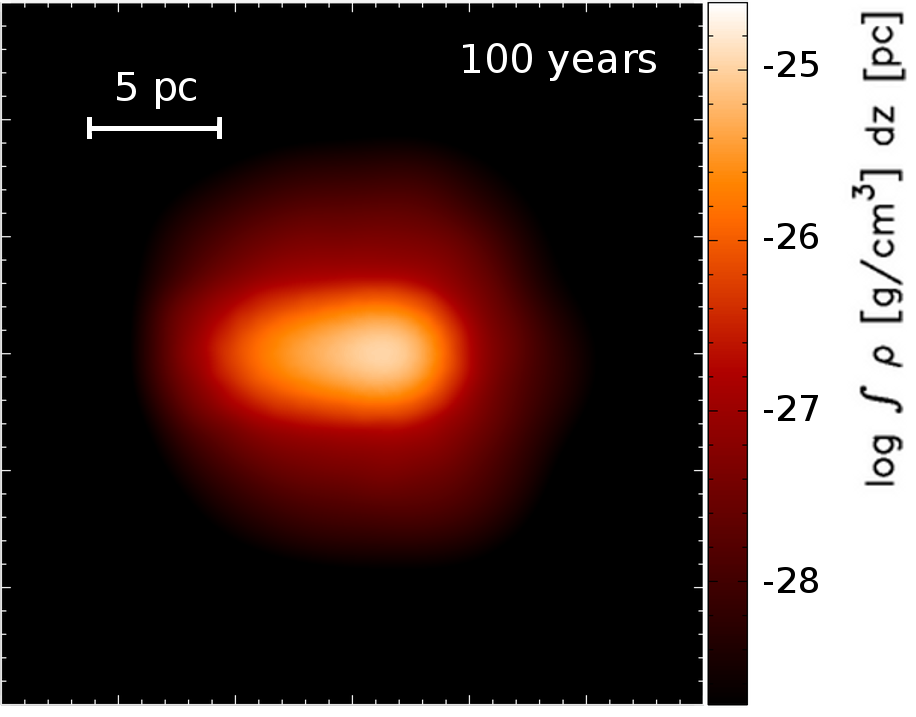}
              }
   \centerline{ \includegraphics[height=3.7cm]{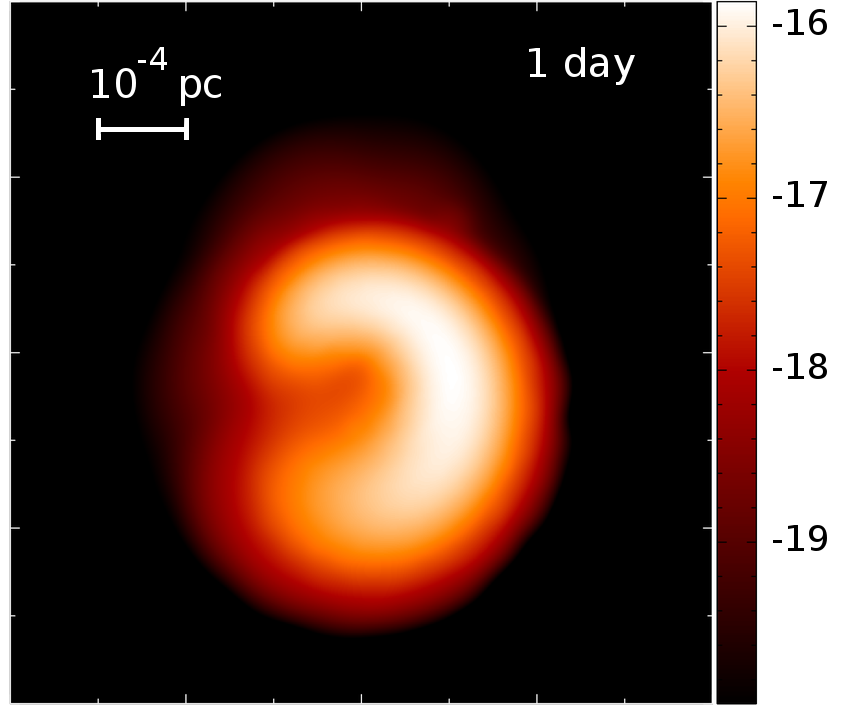} 
                \includegraphics[height=3.7cm]{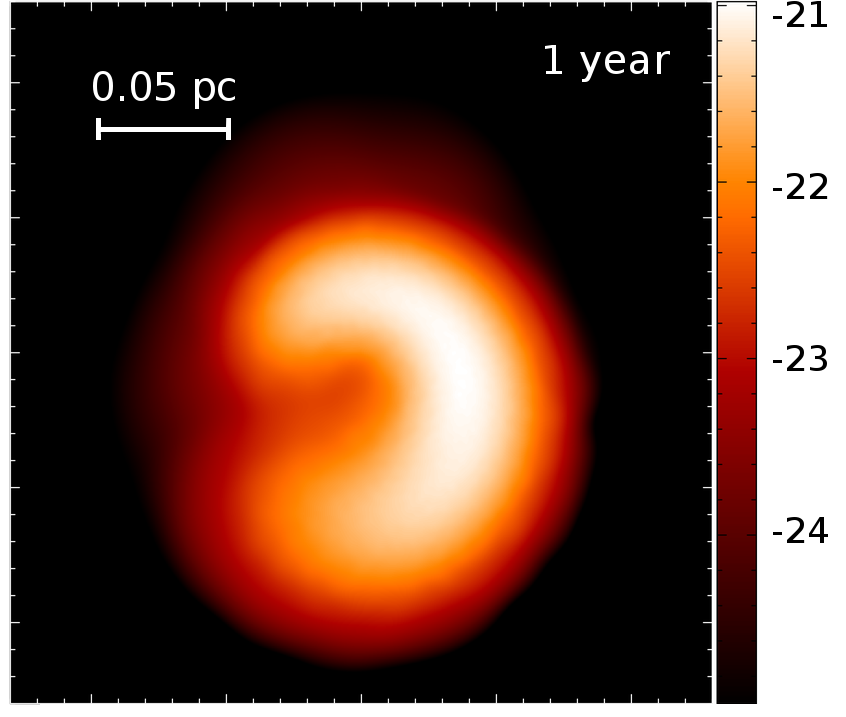} 
                \includegraphics[height=3.7cm]{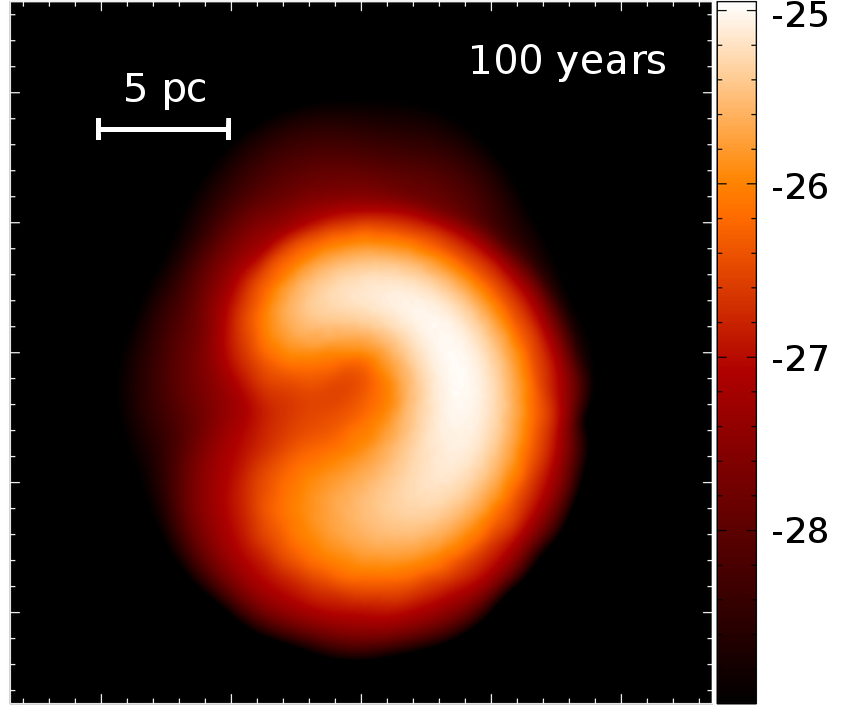}
                \includegraphics[height=3.7cm]{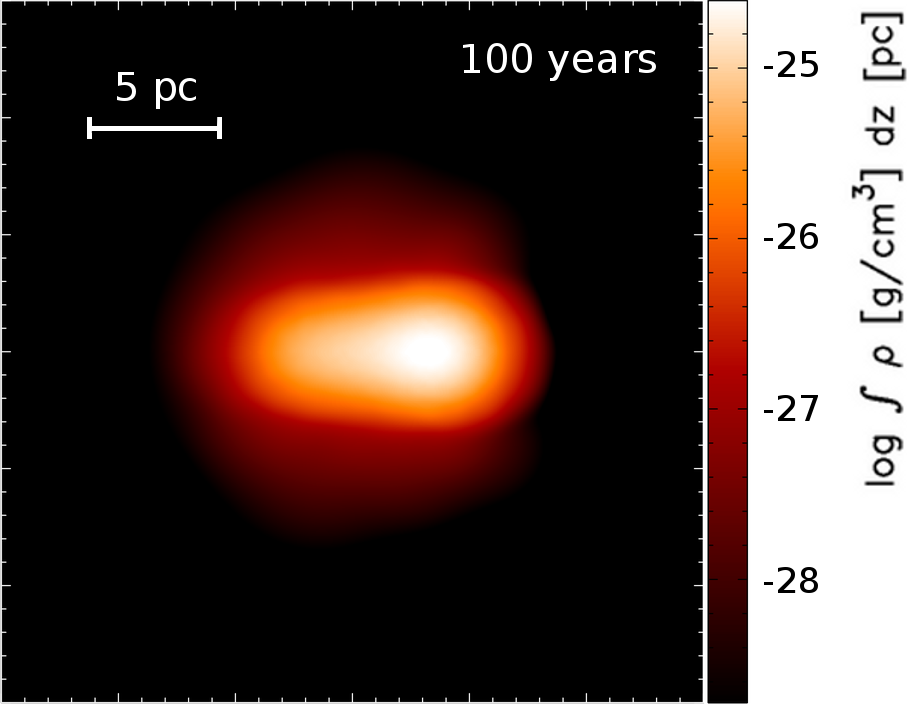}                                                                   
              }
     \caption{Remnant structure at later times. First row: $2 \times 1.4$ \msun,
     second row: 1.4 and 1.3 \msun, third row: 1.6 and 1.2 \msun, last row: 1.8 and 1.2 \msun. For each case the snapshots show
      the column densities after 1 d (first column), 1 year (second column) and after 100 years (third column). The fourth column
      shows the XZ matter distribution at $t= 100$ years for each case.}
   \label{fig:remnant_structure}
\end{figure*}

\subsection{Heating prescription in previous studies}
The rate of nuclear energy generation as a result  of r-process 
nucleosynthesis in the ejecta is a crucial input for macronova models.  It is 
directly reflected in the light curves and hence it has
a decisive impact on their detectability. 
Therefore, it is worth briefly comparing the 
heating rates that have been employed in the different approaches. \cite{tanaka13a} use
an energy generation rate that is based on  the result of running a nuclear reaction 
network \citep{metzger10b} over the thermodynamic trajectory of a single SPH particle 
from an early neutron star merger simulation \citep{rosswog99}.  Since at that time
no electron-/positron-captures were included in the models the electron fractions
in the ejecta were considered as rather uncertain, although by now it has turned out
that the $\beta$-captures only cause minor changes and the ejecta-$Y_e$ stays
close to the initial, cold $\beta$-equilibrium value. For that reason, \citep{metzger10b}
adopted an initial value of $Y_e=0.1$ to determine the energy generation rate. 
\cite{barnes13a} used the heating rates from the work of \cite{roberts11} 
where the initial electron fraction of the ejecta was fixed to a value of $Y_e=0.2$.
Our calculations \citep{rosswog13a,rosswog13b}, in contrast, start from a realistic cold
$\beta$-equilibrium and allow for $Y_e$-changes due to electron-/positron-captures.
We find some trajectories with higher $Y_e$, but the large majority is ejected with a 
very low value of $Y_e \approx 0.03$. While all of these values may appear reasonable,
it actually turns out that the initial value of the electron fraction does matter and  
the lower initial electron fraction leads to a lower energy production at late times.
To illustrate this, we perform a simple experiment: we take the average trajectory 
from our reference case, run B, and calculate the energy generation rate once from our 
best estimate (=0.04) and once with an artificially increased value of $Y_e=0.2$. 
The first is shown as red, and the second one as the dashed green line in 
Fig.~\ref{fig:compare_edot}. Overall there is reasonably good long-term 
agreement as can be seen in  the left panel. However, after a few days, at the time
when the macronovae are expected to peak, the artificially increased $Y_e$-case overproduces the 
heating by a factor of 2--3. 
As can be seen in Fig.~\ref{fig:compare_edot}, such high $Y_e$ also results  
in a less steep decay (with power-law index of 1.2 instead of 1.3). 
The rate from \cite{tanaka13a}, see the blue dotted line, exceeds our heating rate 
by nearly a factor of $4$. These differences will have significant implications on the 
detectability of the macronovae signals, as we discuss in section 6 of Paper II.
For now, we can only speculate that this effect may be due to the
less neutron-rich ejecta producing on average lighter radioactive elements with 
longer half-lives which consequently release their energy later. This
question is left for future work.

\section{Impact of the heating on the remnant structure}
\label{sec:remnant}

To explore the impact that the continuous energy 
injection has on the remnant morphology
we study the evolution of the remnant from a 1.3 and a 1.4 \msunsp
ns (run B in Tab.~\ref{tab:runs}),
once with and once without nuclear energy input (see  Fig.~\ref{fig:heating_vs_no_heating1} left
and right column; as explained above, the innermost matter has been cut out and replaced by a point
mass). 
Total nuclear energy which does not escape in the form of neutrinos can be
roughly estimated as~\citep[cf.][]{metzger10a}:
\begin{align}
E_{\rm nuc}\approx
 \epsilon_{\rm th} X_n ((B/A)_r-\Delta_n)
 \approx 3.6 X_n\;{\rm MeV}\;{\rm nucleon}^{-1},
\end{align}
where 
  $(B/A)_r$~$\approx$~8~${\rm MeV}\;{\rm nucleon}^{-1}$ is an approximate average
  binding energy of the r-process nuclei,
  $\Delta_n\approx0.782$~MeV is the Q-value of neutron decay, and
  $\epsilon_{\rm th}=0.5$ is the adopted value for thermal efficiency (see
  Sec.~\ref{sub:hydrosim}).
The quantity $X_n$ is the initial mass fraction of neutrons, which is
determined from nuclear statistical equilibrium (NSE) and turns out to be
around $X_n\approx0.89$ under typical thermodynamical conditions encountered in
the dynamical ejecta.

Table \ref{tab:runs} lists more accurate estimates of the total retained
nuclear energy $E_{\rm nuc}$, taken directly from the nucleosynthesis network
calculations. It is non-negligible in comparison to the kinetic energy and, indeed,
the continuous energy injection smoothes out initial
inhomogeneities and bloats up the remnant in comparison to the purely adiabatic evolution.
We illustrate the density evolution of the ejecta during the first 100 years by taking the SPH particle
distributions at the data outputs closest to prescribed times ($t_0$: start of longterm evolution, 1 d, 1 month,
100 years) and we show 50 bins of masses and radii in Fig.~\ref{fig:density_evolution}. The average densities drop
within years by 36 (40) orders of magnitude with respect to the longterm simulation start (the initial densities
inside the neutron stars).\\
Again for our reference case with 1.3 and a 1.4 \msunsp 
ns (run B in Tab.~\ref{tab:runs}),
we gauge the impact of the heating on temperature and 
density evolution. We monitor the maximum and average values of density and temperature,
each time with and without heating (see Figs.~\ref{fig:heating_vs_no_heating2_T} and 
\ref{fig:heating_vs_no_heating2_rho}). 

If heating is ignored, both peak (diamond symbol) and average temperatures (dashed line) decrease monotonically due to
$P dV$ work, while the nuclear energy release from radioactive decays leads to a temperature increase up to $\sim 10^9$ K 
after about 0.5 s. For $t \ga 1$ s the temperatures obtained with heating are typically an order of 
magnitude larger than in the case when heating is ignored. The density evolution for both cases 
is qualitatively similar, but at any given point in time the density in the heating case is 
approximately an order of magnitude lower.\\
Fig.~\ref{fig:remnant_structure} shows snapshots
of the remnant column density after 1 d (a typical time for `macronovae' to peak) and also after 1 and 100 years 
(at these times, depending on the ambient medium density and the exact ejecta properties, see equation~(\ref{eq:tau_dec}),
the remnant starts becoming decelerated and should produce radio flares). All remnants expand in a 
nearly perfectly homologous manner. In the symmetric case ($2 \times 1.4$ \msun) matter is `donut-shaped' after 
one day and keeps expanding self-similarly up to 100 years, when it reaches a radius of $r_{\rm cyl} \sim 8$ pc 
and a hight $z\sim 3$ pc. In other words sphericity has not been reached and the remnant still clearly 
remembers the original orbital plane. So 100 years after the merger the remnant still carries the imprint 
of the initial binary mass ratio, equal (two-armed spiral) and unequal mass systems (one-armed spiral) can 
still be clearly distinguished.

\begin{figure}
 \centerline{ \includegraphics[width=0.5\textwidth]{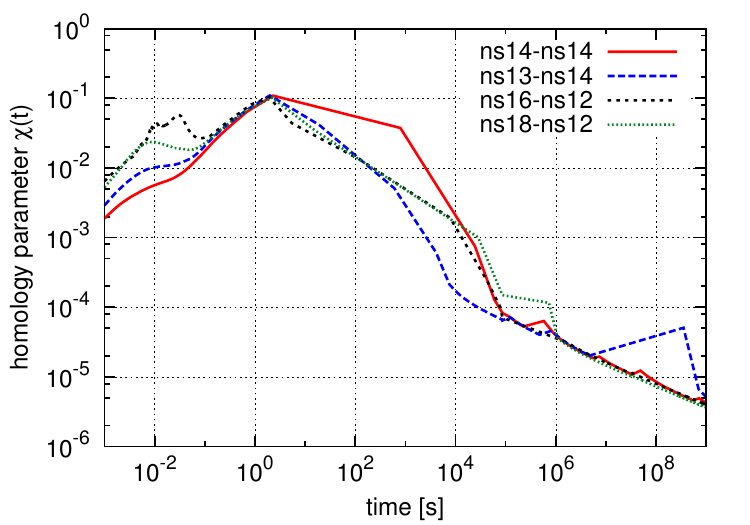}} 
 \caption{ 
  Time evolution of the homology parameter, defined as the relative
  change of velocity due to dynamic acceleration:
  $\chi(t)\equiv\frac{\bar at}{\bar v}$.
  Distributions become homologous up to 1 per cent after about 100~s for all
  non-equal mass cases, and after about 2000~s for the symmetric case of 
  1.4-1.4~\msun.
 } 
\label{fig:homology}
\end{figure}

To quantify the deviations from a perfectly homologous evolution we introduce 
a homology parameter 
\begin{equation}
\chi(t)\equiv\bar at/\bar v
\end{equation}
 with $\bar v$ and $\bar a$ being the 
average velocity and acceleration. In the case of perfectly
homologous expansion, this parameter should be equal to zero.
Fig.~\ref{fig:homology} depicts the evolution of $\chi(t)$ for the considered cases.
After an initial  increase, $\chi(t)$  reaches a maximum at $t\sim1$~s, 
the time when the ejecta have maximal acceleration due to the 
radioactive heating.
After $t= 100$~s the unequal mass cases are homologous to better than
1 per cent, while the equal-mass case of 1.4--1.4~\msunsp reaches the same degree 
only after about 2000~s. At the times that are relevant for our macronova 
calculations ($\sim$1~d), the expansion is homologous to 0.01~per cent in all cases.

\begin{figure*} 
   \centerline{
     \includegraphics[width=9.5cm,angle=0]{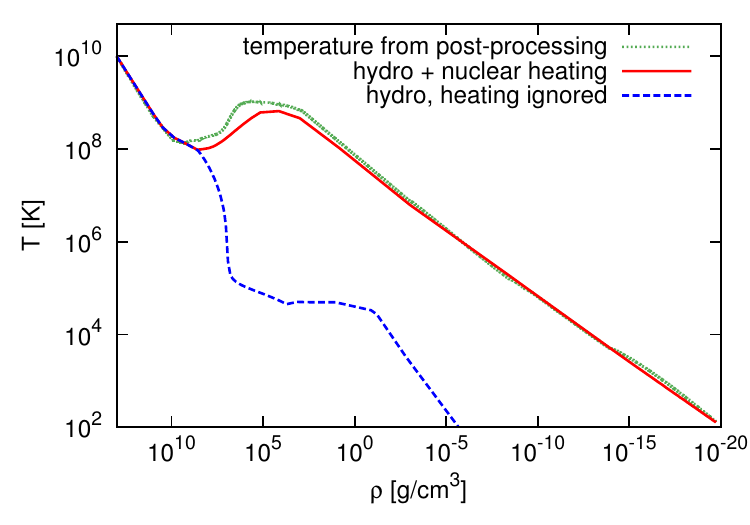} \hspace*{-0.5cm}
     \includegraphics[width=9.5cm,angle=0]{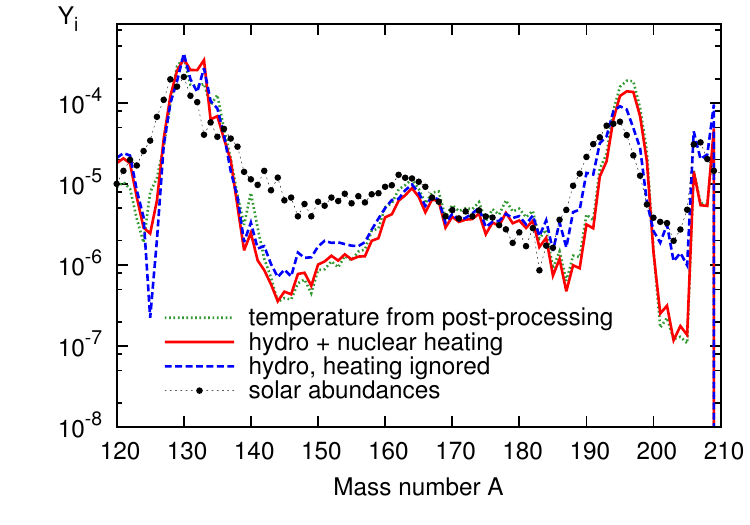} 
   }
      \caption{Assessing the quality of post-processed temperatures. To illustrate the quality of the post-processing 
      approximation (assume density evolution as given obtain the temperature from the entropy production due
      to nuclear reactions) we compare it for a randomly chosen trajectory with the corresponding
      trajectory whose temperature comes from evolving hydrodynamics and nuclear heating concordantly. The blue 
      line labels the `hydrodynamic' temperatures when heating is ignored, and the red line is the result from the `hydrodynamics
      + nuclear heating' calculations. The post-processed temperature is overlaid as the green line. Overall, the agreement
      is remarkably good. The resulting abundances are also in close agreement (`hydrodynamics + nuclear heating' in red,
      `post-processed' in green.}
   \label{fig:orig_vs_postprocessed}
\end{figure*}

\section{Nucleosynthesis}
\label{sec:nucleo}

\subsection{How accurate are nucleosynthesis calculations with post-processed temperatures?}
It is of particular interest to see whether a consistent accounting for the heating in 
the hydrodynamics is crucial for the final r-process abundance distribution. So far, starting 
with \cite{freiburghaus99b} most existing studies have simply 
post-processed existing trajectories with nuclear networks to estimate temperatures, 
thereby implicitly assuming that the accelerated expansion due to the nuclear heating 
is negligible. 
A notable exception is the study by~\cite{goriely11b}, which has studied the effect by means
of a one-zone dynamical model with nuclear heating feedback on the density, therefore accounting for an additional
expansion. However, in their model the observed effect on the density evolution was very small.
By inspecting randomly chosen trajectories from our reference run B 
(with/without heating), we find that the post-processed temperatures are actually 
rather accurate. In Fig.~\ref{fig:orig_vs_postprocessed},
left panel, we compare the different temperature evolutions for the case of an exemplary  
particle trajectory: once with heating ignored and temperature evolution determined 
entirely by adiabatic expansion (dashed, blue, `hydro, heating ignored'), once with the
temperature directly taken from the hydrodynamic evolution including heating (as detailed in
Appendix A; solid, red, `hydro + nuclear heating') and once as reconstructed by running a nuclear
network over a purely hydrodynamic trajectory (as detailed in \cite{freiburghaus99b}; solid, green,
`temperature from post-processing'). This post-processed temperature actually agrees closely with
the one found when heating is accounted for properly. Since the additional expansion due to the
radioactive heating is ignored, the post-processing approach slightly overestimates the temperatures.
Given that the nuclear energy release has a substantial impact on the long-term matter evolution, see below,
it is not self-evident that abundances can be reliably calculated in this ad hoc
manner. Nevertheless, it turns out that the final abundance patterns closely agree with each 
other (see Fig.~\ref{fig:orig_vs_postprocessed}).

\subsection{Dynamic ejecta versus neutrino-driven winds}
We had recently explored the nucleosynthesis inside the dynamic ejecta of compact binary mergers
\citep{korobkin12a} and found that all of the ejecta matter undergoes a very robust, 
`strong' r-process \citep[also confirmed in][]{bauswein13a}. The abundance patterns showed some
sensitivity to the nuclear physics input, but are essentially independent of the parameters of the
binary system: all coalescences produced practically identical abundance patterns beyond $A= 130$, 
it does not even matter whether a nsns or a nsbh system is merging.\\ 
But as pointed out above, compact binary mergers also eject matter via different channels.
With neutrino luminosities of $\sim 10^{53}$ erg/s, neutron star merger remnants
drive strong baryonic winds \citep{ruffert97a,rosswog02b,rosswog03b,rosswog03c,dessart09},
similar to new-born neutron stars \citep{duncan86,qian96b}.
This is an additional mass-loss channel and it can very plausibly complement the heavy 
element nucleosynthesis and produce additional electromagnetic transients.\\
The total amount of mass that is ejected by such winds is not trivial to estimate. \cite{dessart09}
find overall mass-loss rates of $\dot{M}^{\rm wind} \sim 10^{-3}$ \msun/s. Once the central object collapses
to a black hole this rate is expected to drop abruptly since a substantial part of the neutrino emission
comes from its surface layers. For an assumed collapse after 100 ms \cite{dessart09} estimate
$< 10^{-4}$ \msunsp to be ejected by winds. This number, however, is rather uncertain and could be easily
substantially larger for a number of reasons. First, the time scale for the collapse is not well known
and with recent estimates of the minimum, cold neutron star mass around 2.0 \msunsp \citep{antoniadis13}
the differentially rotating central core with temperatures in excess of 10 MeV could be stable for much 
longer. In fact, it is entirely plausible that the low mass end of the neutron star binary population 
could produce a very massive neutron star as final product rather than a black hole. 
The wind would then have a substantially longer duration, comparable to the Kelvin-Helmholtz time scale 
of many seconds. For such cases, however, it remains an open question whether/how baryonic pollution 
could be avoided and a (short) GRB could be launched. Secondly, neutron stars are endowed by possibly strong
initial magnetic fields and the dynamics during a merger offers ample opportunities to amplify 
these initial seeds \citep{price06,liu08,anderson08b,rezzolla11,giacomazzo11,zrake13} to substantial 
fractions of the equipartition strength. A merger remnant rotating at $\sim 1$ ms with a strong magnetic 
field ($\sim 10^{16}$ G) could easily increase the amount of launched mass by orders of magnitude
\citep{thompson03b}.  For these reasons, we parametrize the mass in these winds in
a range from $10^{-4}$ to $10^{-2}$ \msun.\\
In the following we apply a very simple wind model. It is meant to illustrate basic features
of the nucleosynthesis and to discuss the plausibility of a second radioactively powered transient beside the usual
`macronovae' from the dynamic ejecta (see Paper II). This topic deserves more work beyond our simple model,
ideally with multi-dimensional neutrino-hydrodynamic simulations. \\
In our simple approach we calculate the bulk properties of $\nu$-driven winds by means 
of the estimates from \cite{qian96b}. The asymptotic value of the wind electron fraction 
can be estimated as 
\begin{equation}
Y_e^{\rm fin, wind} \approx \left( 1 + \frac{L_{\bar{\nu}_e}}{L_{\nu_e}} 
\frac{\epsilon_{\bar{\nu}_e} - 2 \Delta + 1.2 \Delta^2/\epsilon_{\bar{\nu}_e}}
     {\epsilon_{\nu_e} + 2 \Delta + 1.2 \Delta^2/\epsilon_{\nu_e}} \right)^{-1},
\end{equation}
where $L_{\nu_e}/L_{\bar{\nu}_e}$ are the luminosities of electron neutrinos and anti-neutrinos,
$\epsilon= \langle E^2 \rangle/\langle E \rangle$, $E$ being the neutrino energy, and 
$\Delta$ the neutron-proton mass energy difference of 1.293 MeV. To estimate $\epsilon$ we 
simply multiply our values for $\langle E \rangle$ by a factor of 1.3, as appropriate for 
Maxwell-Boltzmann distributions.
If the neutrino properties from the simulations are inserted (see Table~\ref{tab:runs}),
one finds values between $Y_e^{\rm fin, wind}= 0.28$ and 0.40. 
Interestingly, the asymptotic electron fraction increases with decreasing mass ratio, so that the symmetric 
system produces the lowest $Y_e$ values. Again based on \cite{qian96}, we find that the average 
entropies in the wind are very close to 8 $k_B$ per baryon for all our cases.\\
In our simple model, we produce a synthetic trajectory with a linear expansion
  profile $\rho(t) = \rho_0 (1 + vt/R_0)^{-3}$, starting from the entropy 
  $s_0 = 8\;k_B/{\rm baryon}$ and a range of electron fractions,
  corresponding to the runs A-D.
  Based on the results of \cite{dessart09}, their fig.~2, we select the initial 
  density $\rho_0=5\times10^7\;{\rm g}{\rm cm}^{-3}$ and the characteristic radius 
  $R_0=200\;{\rm km}$ in a way that temperature is safely above the NSE
  threshold.
  We then run the nucleosynthesis network, using the calculated values of
  $\rho(t)$ and self-consistently incrementing the entropy in the same way as
  described above for the dynamical ejecta case.
  The temperature of the trajectory at every time is calculated using the Helmholtz
  EOS and the needed values of $\bar{A}$ and 
  $\bar{Z}$ are computed from the network. For the expansion velocity, we use 
  the escape velocity from the launch region ($v = 0.11c$).\\
\begin{figure}
  \begin{center}
    \begin{tabular}{cc} 
    \includegraphics[width=0.48\textwidth]{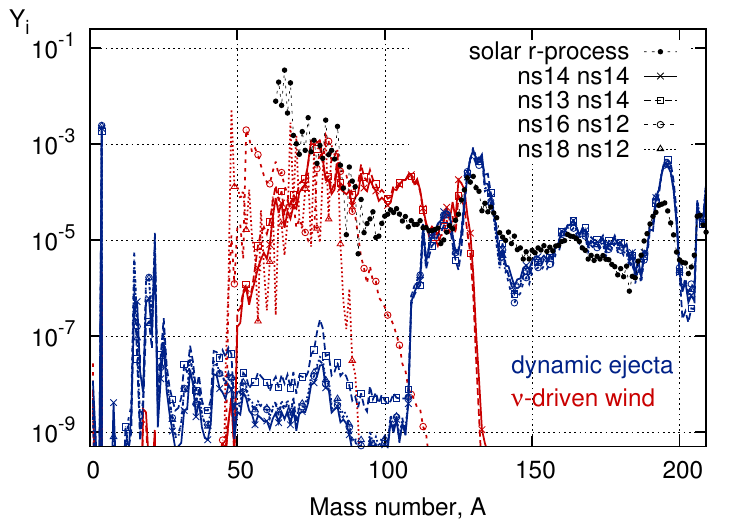} 
    \end{tabular}
    \caption{Robustness versus variability in the abundance patterns:
             final abundances from both the dynamic ejecta (blue) and the neutrino-driven wind (red) for our
             four simulations (each of the abundance curves is normalized separately, $\sum_i A_i Y_i=1$).
             The numbers in the 
             legend indicate the neutron star masses in units of 0.1 \msun, i.e. ns13 ns14 refers to a system with 
             a 1.3 and a 1.4 \msunsp neutron star. Note that dynamic ejecta produce an extremely robust abundance pattern
             for $A > 110$, while the patterns in  neutrino-driven winds vary strongly between different merger events.}
    \label{fig:robust_vs_variable}
  \end{center}
\end{figure}
\begin{figure*}
  \begin{center}
    \hspace*{0cm}\includegraphics[width=0.95\textwidth]{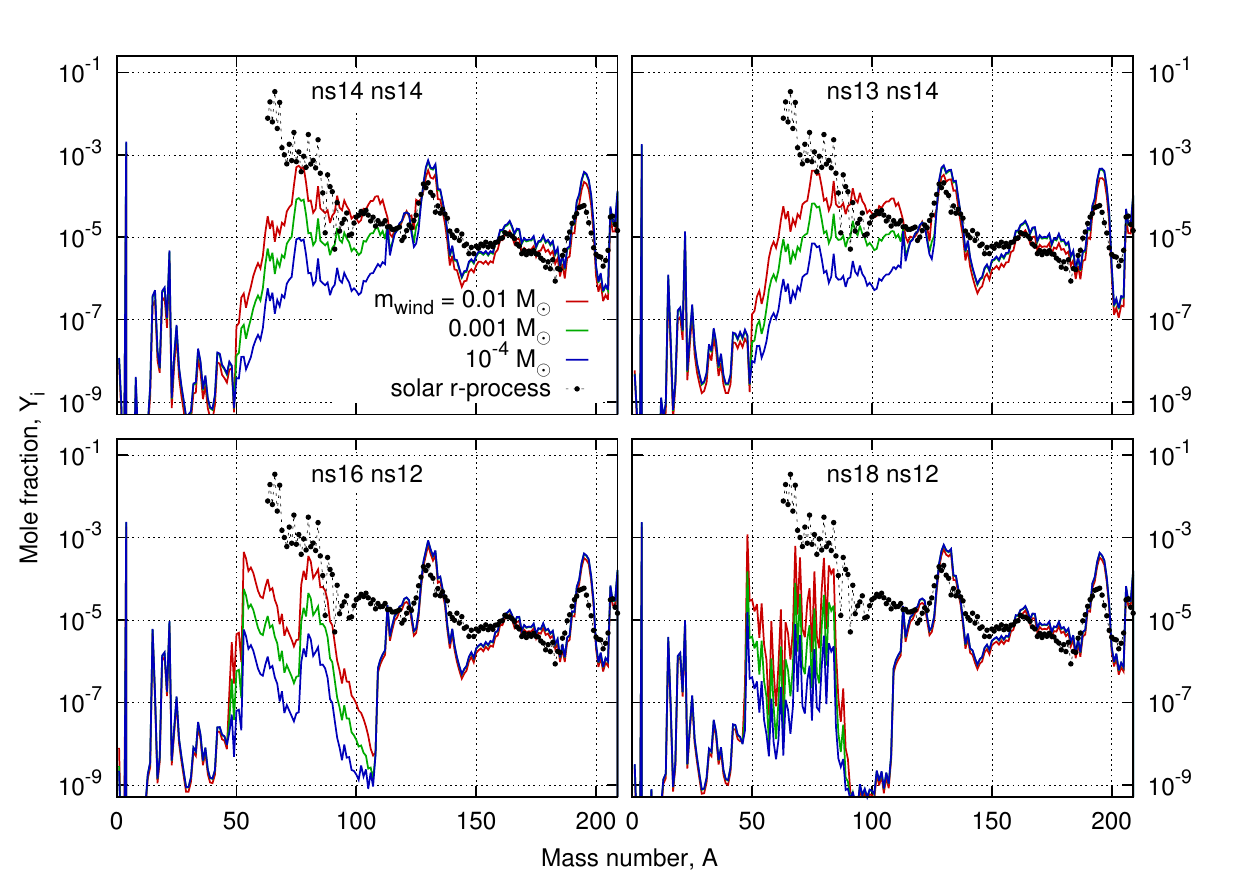}
    \caption{Resulting abundances from the sum of the dynamic ejecta and the neutrino-driven winds for 
             all four models. For the dynamic ejecta we use the masses ($m_{\rm ej}$) as found in the simulations, see Tab.~\ref{tab:runs},
             for the neutrino-driven winds the masses are poorly known and therefore parametrized from $m_{\rm wind}= 10^{-4} ...  10^{-2}$
             \msun. Note that the abundance pattern of the wind component varies substantially between different runs, while
             it hardly varies for the dynamic ejecta component ($A\ga 110$).}
    \label{fig:AbundancesDynejWinds}
  \end{center}
\end{figure*}
We find element distributions between $A=50$ and 130 in a weak r-process. 
In Fig.~\ref{fig:robust_vs_variable} we show the resulting nucleosynthesis for all four runs
and each time we distinguish between the dynamic ejecta (blue) and the $\nu$-wind (red). 
Note that this shows the individual compositions for these two types of ejecta, not 
considering how much ejecta mass is involved in each of them.
As pointed out by \cite{korobkin12a}, the dynamic ejecta abundances for $A \ga 130$ hardly vary
at all from case to case. The $\nu$-wind abundances, in contrast, produce matter in the range from 
$A=50$ to 130 with substantial variations between different merger cases. The latter is due 
to the sensitivity of the nucleosynthesis to $Y_e$, which, in turn, is set by the neutrino properties.
In Fig.~\ref{fig:AbundancesDynejWinds} we show the abundances for both wind and dynamic ejecta. We use
the simulation results for the masses of the dynamic ejecta and  $10^{-4}$, $10^{-3}$ and 
$10^{-2}$ \msunsp for the masses in the wind. The differences in the abundances below the second peak 
will naturally lead to variations in the relative abundances between heavy r-process elements and elements 
with $A<130$, and this is important to understand the role of neutron star mergers in 
the chemical history of our galaxy. \\
We also note that the neutrino-driven winds produce a range of radioactive isotopes (though no $^{56}$Ni)
that are long-lived enough to still be present when the wind matter becomes transparent, i.e. one can expect a 
second, radioactively powered transient. Its details are discussed in Sec. 5 of Paper II. 
It turns out that our simple neutrino-driven wind model produces an electromagnetic transient
that is more promising for coincident detection with a GW signal than the `macronova'
signal from the dynamic ejecta, mainly due to the different opacities
\citep{kasliwal13}.

\section{Summary and discussion}
\label{sec:discussion}
In this work we have focused on the long-term evolution of the dynamic ejecta of a neutron star merger
and in particular on the role that the freshly synthesized and radioactively decaying r-process elements
play. We have included the energetic feedback from radioactive decays based on nuclear network
calculations into the hydrodynamic evolution. Contrary to existing simulations which typically stop after
$\sim$ 20 ms, we follow the ejecta evolution through all phases where electromagnetic
emission is expected to occur and we only stop the simulations after 100 years. At this time
(at latest) the ejecta should have swept up ambient matter comparable to their own mass and start becoming 
decelerated, a process that is not modelled here.
These simulations allow us to accurately quantify how homologous the expansion actually is. We find that in all cases the degree 
of homology after $10^4$ s is better than 0.1 per cent (see Fig.~\ref{fig:homology}).
\\
We find that the radioactive heating has a substantial impact on the morphology and efficiently 
smoothes out initial inhomogeneities.
Although the nuclear energy input does alter dynamics and morphology it does not erase the memory 
of the initial binary system parameters. For example, the remnant matter  keeps its `donut shape'
in the case of a perfectly symmetric ($q=1$) merger until at least 100 years after the coalescence,
while asymmetric mergers still carry the imprint of their initial mass ratio. \\
\begin{figure}
  \begin{center}
    \begin{tabular}{cc} 
    \includegraphics[width=0.48\textwidth]{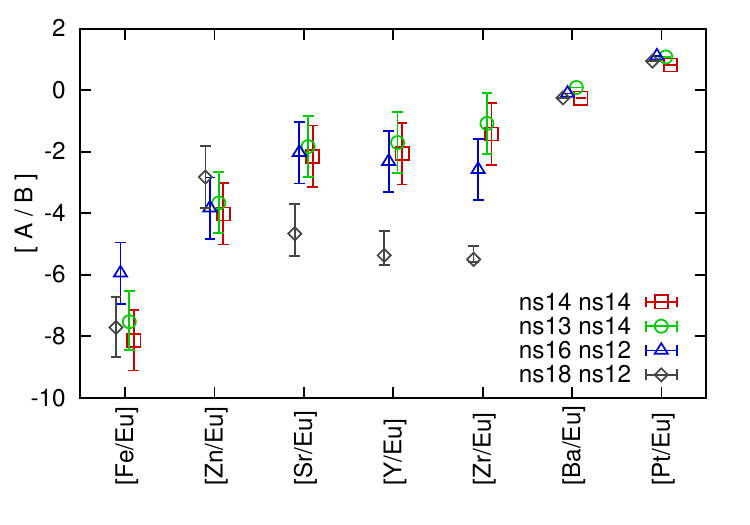} 
    \end{tabular}
    \caption{Abundance ratios of representative elements with respect to Eu for the sum of dynamic ejecta and 
               neutrino-driven winds. The error bars reflect the estimated uncertainty in the mass ejected in 
               neutrino-driven winds, which is varied from $10^{-4}$~\msunsp and $10^{-2}$~\msun. Ba and Pt represent 
               strong r-process components and exhibit very little scatter with respect to both the merging system, 
               and the unknown mass of the wind component. Sr, Y and Zr are the weak r-process elements, which 
               are very sensitive to the merging system, in particular the electron fraction. Fe and Zn represent 
               the iron group elements; they are under-produced for all combinations of masses and merging systems.}
    \label{fig:abundance_ratios}
  \end{center}
\end{figure}
A number of recent studies [\cite{panov08}, \cite*{goriely11a}, \cite{goriely11b,roberts11,korobkin12a,bauswein13a}]
have dealt with the dynamic ejecta of neutron star mergers. 
Most studies have simply post-processed the temperature along trajectories of prescribed density,
thereby ignoring the effect of the radioactive heating on the density evolution. Only ~\cite{goriely11b} have
investigated this effect by means of a  one-zone expansion model. None of the studies that we are 
aware of has included the energetic feedback from nuclear reactions in a 3D hydrodynamic simulation.
But despite the large impact in the long-run, 
we find that such post-processed nucleosynthesis results agree
actually well with those calculated including nuclear
heating. This is mainly because in the early stages where the r-process takes place (seconds), the impact of the
energy release on the density evolution is still small.\\
While each of the above mentioned studies had their specific focus and contribution, they all agree
with the basic picture outlined in \cite{freiburghaus99b}: the dynamic ejecta seem to naturally produce
the `strong' r-process component  and the ejected amounts of matter are
consistent with contributing a major part to the overall galactic r-process inventory.
The resulting
abundance pattern is essentially independent of the exact system that merges \citep{korobkin12a}, i.e.
it hardly matters whether two neutron stars of different masses or even a neutron star black hole system
merges.  All these events produce essentially identical abundance signatures above $A \ga 130$, this is
illustrated for runs A-D in Fig.~\ref{fig:robust_vs_variable} (blue curves).\\
For two reasons we consider the conclusion that the dynamic ejecta are excellent candidates for 
the `strong' r-process as very robust. First, the ejecta amounts found by different 
groups are not identical, but given the large uncertainties in the merger rates all of them are consistent 
with being an important r-process source. For nsns mergers we found a range from 
$8 \times 10^{-3}$ - $4 \times 10^{-2}$ \msunsp \citep{rosswog13b}, approximate GR calculations find  
$10^{-3} - 2 \times 10^{-2}$ \msunsp \citep{bauswein13a} and full GR calculations \citep{hotokezaka13} 
find $10^{-4} -  10^{-2}$ \msun. Even the results for our Newtonian nsbh calculations agree quite well 
with the GR results [see Tab. 1 in \citep{rosswog13b} and \cite{kyutoku13}]. Secondly, the robustness of
the abundance pattern is due to the extreme neutron-richness of the ejecta, we find typical values of
$Y_e\approx 0.03$ \citep{korobkin12a,rosswog13a}. This value is determined by the cold $\beta$-equilibrium
in the initial neutron star and it could change for stars of different compactness (say, due to a different
EOS or a different treatment of gravity). But in this range even $Y_e$-variations by factors
of a few hardly change the resulting abundance pattern. Only at values beyond $Y_e\approx 0.18$ does the 
pattern of the heaviest elements ($A \ga 130$) become sensitive to the exact value of $Y_e$ (see fig.~8 
in \cite{korobkin12a}).
This makes compact binary mergers natural candidates for the sources of the robust, `strong' r-process 
component. Whether neutron star mergers can also produce the earliest  enrichments of galaxies
with r-process elements remains to be further investigated, though.\\
We have also briefly investigated the nucleosynthesis in the neutrino-driven winds from a neutron star
merger remnant. The nucleosynthesis therein is mainly determined by the asymptotic electron fraction $Y_e$, which, in
turn, is set by the neutrino properties. These winds produce a weak r-process with elements from $A=50$ to
130 and the abundance patterns vary substantially between different merging systems. 
We stress once more that the model discussed here is very simple, but it indicates that this topic 
is worth more detailed neutrino-hydrodynamics studies.\\
Observations of low-metallicity stars give an indication that the r-process elements related to the production
of the heaviest elements (an indicator is the r-process element Eu) are not correlated with the Fe-group, 
i.e. Fe is not co-produced in events responsible for the Eu production, or provides at most a minor 
contribution \citep{cowan05}. Intermediate-heavy elements like Sr, Y, Zr (also called LEPP elements)
show a weak correlation \citep{cowan05}, i.e. can be co-produced, but probably have their dominant 
contribution from a different process ($\nu$-p process, weak r-process, charged particle
process, see e.g. \cite{qian07,qian08}). The co-production with Eu for those stars most 
strongly enriched in this heavy r-process element, seems to give a contribution of about 5--10 per cent to the total Sr,
Y, Zr and Ag  production (see fig.~1 in \cite{montes07}). On the other hand,  Ba, La, Ce, Nd
and Sm, as signatures of a strong r-process, are all co-produced in their solar r-process contribution.\\
If neutron star mergers are the source of this heavy element r-process, we expect a similar behaviour in their
ejecta. Fig.~\ref{fig:abundance_ratios} gives an indication of exactly this behaviour with a Ba/Eu ratio 
equaling the solar r-process pattern, while Sr, Y, Zr ratios are down by a factor of 10-100, and the 
Fe-group shows a negligible contribution. Note that in (total) solar abundances
Ba is dominated by the s-process. This explains the difference between [Ba/Eu] and 
[Pt/Eu] in Fig.~\ref{fig:abundance_ratios}, which in dynamic merger ejecta are produced in their 
solar `r-process contribution'
(see Fig.~\ref{fig:AbundancesDynejWinds}). Thus, even for the highest possible contribution of 
wind ejecta, weak or strong correlations are not destroyed and neutron star mergers provide exactly the 
behaviour required from the strong r-process source.

\section*{Acknowledgements}

We would like to thank C. Winteler for providing his nucleosynthesis network
code and for his continued support.
This work has also benefited from the stimulating discussions at
the MICRA workshop in 2013.
This work was supported by DFG grant RO-3399, AOBJ-584282 and by the Swedish Research Council (VR) 
under grant 621-2012-4870. The simulations of this paper were performed on the facilities of the
H\"ochstleistungsrechenzentrum Nord (HLRN). A.A. is supported by the Helmholtz-University 
Young Investigator grant No. VH-NG-825. F.-K. T. gratefully acknowledges support
from the Swiss National Science Foundation (SNF), both S.R and F.-K. T. have been supported
by Compstar. A.A. and F.-K. T. are part of Collaborative Research Program Eurogenesis/MASCHE
funded by the European Science Foundation.
T.P. is supported by an ERC advanced grant (GRBs) and by the  I-CORE 
Program of the Planning and Budgeting Committee and The Israel Science
Foundation (grant No 1829/12).\\

\hyphenation{Post-Script Sprin-ger}

\hyphenation{Post-Script Sprin-ger}\hyphenation{Post-Script Sprin-ger}

\bsp

\appendix

\section{Fit formulae for the heating due to nuclear reactions}
The call of the Helmholtz equation of state requires 
\begin{eqnarray}
\bar{A} &=& \left(\sum_i \frac{X_i}{A_i}\right)^{-1}\label{eq:Abar}\\
\bar{Z} &=& \bar{A} \sum_i Y_i Z_i\label{eq:Zbar}
\end{eqnarray}
on input, where the index runs over all nuclear species including neutrons and protons.
From our study \citep{korobkin12a} we have deduced the following fit formulae:
\begin{eqnarray}
\hspace*{-0.4cm}\bar{A}(t)\hspace*{-0.2cm}&=&   \hspace*{-0.2cm}\left\{
    \begin{array}{ l l l}
     \hspace*{-0.3cm}A_0 + A_1(t_0-t)^{\alpha_1}\quad \quad \quad \quad \quad \quad \quad \quad \quad \; \; \; t \le t_0\\
     \hspace*{-0.3cm}A_\infty + A_2(t-t_1)^{\alpha_2} + C_1  e^{-\frac{t}{\tau_1}} + C_2 e^{-\frac{t}{\tau_2}} \;      t > t_0  
       \end{array} \right.\\
\hspace*{-0.4cm}\bar{Z} (t)\hspace*{-0.2cm}&=&  \hspace*{-0.2cm} \left\{
    \begin{array}{ l l l}
     \hspace*{-0.3cm}Z_0 + B_1(t_0-t)^{\beta_1}\quad \quad  \quad \quad \quad \quad \quad \quad \quad \quad  t \le t_0\\
     \hspace*{-0.3cm}Z_\infty + B_2 t^{\beta_2} + B_3  t^{\beta_3}  \quad \quad \quad \quad \quad \quad \quad \quad    t > t_0  
       \end{array} \right.
\end{eqnarray}
and 
\begin{equation}
 \dot{\epsilon}(t) = 2 \times 10^{18} \frac{\rm erg}{\rm g\cdot s}\left(\frac{1}{2}-\frac{1}{\pi}
                  \arctan{\frac{t-1.3\;{\rm s}}{0.11\;{\rm s}}}\right)^{1.3}
                  \times\left(\frac{\epsilon_{th}}{0.5}\right)
\end{equation}
with the fit parameters having the following values: $t_0= 1.07$ s, $t_1$= 0.4 s, $\tau_1= 7.83 \times 10^{8}$ s, $\tau_2=1.12 \times 10^{5}$ s,
                $A_0= 0.69$, $A_1= 0.4922$, $\alpha_1= -1.328$,
                $A_\infty= 149.5000$, $A_2= 16.7200$, $\alpha_2= -1.9100$, $C_1= 2.1370$, $C_2=0.6838$, 
                $Z_0=-0.2100$, $B_1= 0.2783$, $\beta_1=-1.5373$, $Z_\infty= 61.2000$, $B_2= -3.0000$, 
                $\beta_2=-0.1500$,$B_3=8.8650$, $\beta_3= -3.7970$.

\label{lastpage}

\end{document}